\documentclass[12pt,eadjoint tfnpsf]{article}

\catcode`\@=11
\@addtoreset{equation}{section}

\usepackage{amsbsy,amssymb,latexsym,amsfonts, amsmath}
\usepackage{mathrsfs}
\usepackage{graphicx}

\RequirePackage[dvips,usenames]{color}
\definecolor{fireblick}{rgb}{0.698039,0.133333,0.133333}

\newcommand{\beq}{\begin{equation}}
\newcommand{\eeq}{\end{equation}}
\newcommand{\bea}{\begin{eqnarray}}
\newcommand{\eea}{\end{eqnarray}}

\newcommand{\w}{\wedge}

\newcommand{\CN}{{\mathcal N}}

\newcommand{\CT}{{\mathcal T}}

\def\Tr{\mathop{\rm Tr}}
\newcommand\tr{\mathrm{tr}}

\def\tr{{\,\mathrm{tr}\,}}

\begin{document}
%
%
\begin{titlepage}
\begin{flushright}
\normalsize
~~~~
YITP-09-40\\
July, 2009 \\
\end{flushright}

\vspace{72pt}

\begin{center}
{\LARGE New Seiberg Dualities from $\CN=2$ Dualities}\\
\end{center}

\vspace{22pt}

\begin{center}
{%
Kazunobu Maruyoshi\footnote{e-mail: maruyosh@yukawa.kyoto-u.ac.jp}, 
Masato Taki\footnote{e-mail: taki@yukawa.kyoto-u.ac.jp},
Seiji Terashima\footnote{e-mail: terasima@yukawa.kyoto-u.ac.jp} and
Futoshi Yagi\footnote{e-mail: futoshi@yukawa.kyoto-u.ac.jp}
}\\
%
\vspace{15pt}
%
\it Yukawa Institute for Theoretical Physics, Kyoto University, Kyoto 606-8502, Japan\\
\end{center}
%
\vspace{20pt}
\begin{center}
Abstract\\
\end{center}
  We propose a number of new Seiberg dualities of $\mathcal{N} =1$ quiver gauge theories.
  The new Seiberg dualities originate in new S-dualities of $\mathcal{N}=2$ superconformal field theories 
  recently proposed by Gaiotto.
  $\mathcal{N} =2$ S-dual theories deformed by suitable mass terms flow to our $\mathcal{N} =1$ Seiberg dual theories.
  We show that the number of exactly marginal operators is universal for these Seiberg dual theories
  and the 't Hooft anomaly matching holds for these theories.
  These provide strong evidence for the new Seiberg dualities.
  Furthermore, we study in detail the Klebanov-Witten type theory and its dual as a concrete example.
  We show that chiral operators and their non-linear relations match between these theories.
  These arguments also give non-trivial consistency checks for our proposal.


\vfill

\setcounter{footnote}{0}
\renewcommand{\thefootnote}{\arabic{footnote}}

\end{titlepage}

\tableofcontents

\section{Introduction}
\label{sec:intro}
  Seiberg duality \cite{Seiberg} plays an important role 
  for understanding phase structures of ${\cal N}=1$ supersymmetric gauge theories in four dimensions.
  Although an ${\cal N}=1$ theory and its Seiberg dual theory are not equivalent,
  the dual theory describes the same infrared physics as that of the original theory. 
  For example, in $\CN=1$ supersymmetric QCD (SQCD) in conformal window ($3N_c/2 \leq N_f \leq 3 N_c$), 
  the Seiberg duality implies the existence of a non-trivial infrared fixed point,
  where an interacting superconformal field theory is realized \cite{Seiberg} (see \cite{ISreview} for a review).
  The original and dual theories flow to the same infrared fixed point.
  
  On the other hand, some $\CN=2$ superconformal gauge theories are known to have an exact S-duality,
  which means that a strong gauge coupling region of a theory is equivalent to a weak
  coupling region of another theory at any energy scale \cite{SW, APShapere},
  like Montonen-Olive duality in the $\CN=4$ supersymmetric gauge theory.
  Interestingly, it was proposed in \cite{LS} that the Seiberg duality is associated with the S-duality
  in the ${\cal N}=2$ supersymmetric gauge theory.
  By the mass deformation for the adjoint chiral multiplet, 
  the S-dual pair of $\CN=2$ superconformal gauge theories flows to the $\CN=1$ theories 
  which are precisely a Seiberg dual pair.
  (See \cite{APS, HMS, AILS} for related discussions.)
  
  Recently, Gaiotto proposed a new chain of S-dualities in $\CN=2$ superconformal quiver gauge theories \cite{Gaiotto}
  and many related developments have been made \cite{Gaiotto:2009gz, Tachikawa, BBT, AGT, ND, Wyllard}. 
  The $\CN=2$ superconformal theories associated with the generalized quiver diagrams, 
  which we will explain later, are all equivalent, or S-dual to each other, 
  if they have the same ``genus" of the quiver diagram and the same global symmetry.
  The theories where the gauge group is $SU(2)^{p}$ are the simplest case.
  In this case, we can explicitly construct the Lagrangians of all the generalized quiver gauge theories and 
  their flavor symmetries are generically $SU(2)^n$.\footnote{
    For some $SU(N)$ generalized quiver gauge theories ($N > 2$), the Lagrangian description has not been found.}
  They form a large class of the quiver gauge theories, which is denoted as $\CT_{g, n}$, where $3-3g=n-p$, 
  because they are proposed to realized on M5 branes wrapped on genus $g$ Riemann surface with $n$ punctures. 
  
  In this paper, from the Gaiotto's S-dualities in $\CT_{g, n}$,
  we propose a number of new Seiberg dualities of $\mathcal{N} =1$ quiver gauge theories, 
  which implies there are a large number of new non-trivial $\mathcal{N} =1$ superconformal field theories.
  As one gauge group case in \cite{LS}, 
  $\CN=2$ superconformal quiver gauge theories $\CT_{g, n}$ are expected to flow to infrared fixed points,
  by the mass deformations for the adjoint chiral multiplets.
  This can be partially verified by turning off the gauge couplings except one.
  In that case, the theory is nothing but $\CN=1$ SQCD with four flavors 
  which flows to the non-trivial infrared fixed point.
  
  In general, this deformation produces several quartic terms in the superpotential.
  In the ultraviolet, these are irrelevant operators.
  However, they show non-trivial behavior in the infrared: 
  some combinations of these develop to exactly marginal operators,
  whose coupling constants span a manifold of fixed points.
  It is generally difficult to identify the exactly marginal operators.
  However, we can still count the (complex) dimension of the manifold of the fixed points, 
  that is the number of the exactly marginal operators by means of the argument in \cite{LS}.
  It reveals that if we concentrate on the operators keeping the flavor symmetry $SU(2)^n$,
  the number of them is $2n$ and universal for the quiver gauge theories 
  obtained from $\CT_{g, n}$ for fixed $g$ and $n$.
  
  We focus on the above-mentioned fact that 
  the S-duality relates various different looking $\CN=2$ superconformal quiver gauge theories.
  We show that this property implies interesting physics:
  many different looking $\CN=1$ quiver gauge theories flow to the same infrared fixed point, 
  by the mass deformations of the S-dual family.
  As a result, we propose new Seiberg dualities which relate a large number of $\CN=1$ quiver gauge theories.
  For instance, two quiver theories obtained from $\CT_{1, 2}$ theories are expected to be dual.
  One quiver in this category is a slightly generalized theory of the one considered by Klebanov and Witten \cite{KW}, 
  i.e., the quiver with a loop and two gauge groups.
  The S-dualities imply that this theory is Seiberg dual to the other theory with a different quiver diagram.
  Notice that the original generalized Klebanov-Witten theory is self-dual under the usual Seiberg duality on one gauge group.
  Therefore, it is very remarkable that the new Seiberg dual theory describes the same infrared physics 
  as that of the generalized Klebanov-Witten theory.
  
  In order to verify our proposal, we give non-trivial consistency checks of these dualities.
  The above counting of the exactly marginal operators could be strong evidence:
  recall that it is universal for fixed $g$ and $n$.
  Also, we show that the 't Hooft anomaly matching holds 
  for all the theories obtained from $\CT_{g, n}$ for fixed $g$ and $n$, 
  although we need a bit care in the case with an enhanced global symmetry.
  Another non-trivial check of these dualities is the matching of several operators.
  We will perform this in the quiver gauge theories obtained from $\CT_{1, 2}$ theories mentioned above.
  We demonstrate that chiral operators match between the generalized Klebanov-Witten theory and its dual. 
  We also consider the matching of non-linear relations for the chiral
  operators, i.e., the matching of the chiral ring,
  which indicates that the matching of the classical moduli space. 
  
  The organization of this paper is as follows. 
  After reviewing the S-dualities in $\CN=2$ superconformal quiver gauge theories \cite{Gaiotto} in subsection 2.1,
  we then consider the deformation to $\CN=1$ in subsection 2.2.
  We will propose new Seiberg dualities among these theories.
  Also, we check that the global anomalies match in these theories.
  In section 3, we analyze exactly marginal operators in these theories.
  We will see that the number of the exactly marginal operators is universal for the proposed Seiberg dual theories.
  In section 4, we consider a simple model, whose quiver diagram is identical to 
  the Klebanov-Witten theory, and its dual.
  We will identify the operators in the dual theory, which correspond to the operators in the generalized Klebanov-Witten theory.
  Section 5 is devoted to conclusion and discussion.

\section{$\CN=1$ SCFTs from $\CN=2$ SCFTs}

\subsection{$\CN=2$ superconformal quiver gauge theories and S-dualities}
\label{subsec:Gaiotto}
  A large class of $\CN=2$ superconformal quiver gauge theories in four dimensions was constructed in \cite{Gaiotto}.   
  In this paper, we mainly consider the theories with $SU(2)^p$ gauge group.
  In this case, a class of $\CN=2$ superconformal quivers is specified by 
  the number of $SU(2)$ flavor symmetries, $n$, and ``genus'' of the quiver diagram, 
  $g = 1 + \frac{p - n}{3}$, which is denoted as $\CT_{g,n}$.
  Since there exist many Lagrangian descriptions associated with a label $(g,n)$, 
  quivers contained in $\CT_{g,n}$ are various in shape.
  The different looking quivers with same $g$ and $n$ are related by the S-duality.
  We will review this in this subsection.
  
\begin{figure}
    \begin{center}
    \includegraphics[scale=0.6]{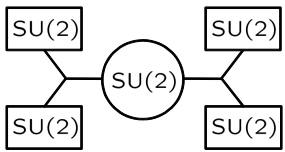}
    \caption{{\small The quiver diagram of $SU(2)$ gauge theory with four fundamental hypermultiplets.}}
    \label{fig:T40}
    \end{center}
    \end{figure}

  Among these $\CN=2$ superconformal gauge theories, 
  the $SU(2)$ gauge theory with 4 fundamental hypermultiplets, $\CT_{0,4}$, is the simplest and important example.
  Since the fundamental representation of $SU(2)$ is pseudo-real, the flavor symmetry is $SO(8)$.
  We consider  an $SO(4) \times SO(4) \simeq SU(2)^4$ subgroup of the $SO(8)$ flavor symmetry according to \cite{Gaiotto}.
  Motivated by this, we denote this theory as Fig.~\ref{fig:T40}.
  
  In order to construct the Lagrangian of this theory, we start with four free fundamental hypermultiplets, 
  which we will denote as $\CT_{0,3}$.
  Let us denote these superfields by $Q^{\alpha a i}$, 
  where three indices label fundamental representations
  of the different flavor $SU(2)$'s ($\alpha = 1, 2$, $a = 1, 2$ and $i = 1, 2$).
  By gauging one of three $SU(2)$ flavor symmetries, e.g., $SU(2)$ labeled by $\alpha$, 
  the superpotential can be written as
    \bea
    W
     =     Q^{\alpha a i} (\varepsilon \phi)_{\alpha \beta} \varepsilon_{ab} \varepsilon_{ij} Q^{\beta b j}.
    \eea
  Now we introduce two copies of four free hypermultiplets to construct $\CT_{0,4}$ theory.
  The $SU(2)$ gauge group can be obtained 
  by gauging a diagonal part of two $SU(2)$ symmetries of different sets of four free hypermultiplets.
  
  In order to consider the S-duality, we introduce the mass parameters associated 
  with $SU(2)_{a, b, \ldots}$ flavor symmetries as $m_{a, b, \ldots}$.
  The S-duality of this theory is associated with the triality of $SO(8)$ which
  exchanges $\textbf{8}_v$, $\textbf{8}_s$ and $\textbf{8}_c$ representations of $SO(8)$.
  Under the $SU(2)_a \times SU(2)_b \times SU(2)_c \times SU(2)_d$ subgroup, they decompose as
    \bea
    \textbf{8}_v
    &=&    (\textbf{2}_a \otimes \textbf{2}_b) \oplus (\textbf{2}_c \otimes \textbf{2}_d),
           \nonumber \\
    \textbf{8}_s
    &=&    (\textbf{2}_a \otimes \textbf{2}_c) \oplus (\textbf{2}_b \otimes \textbf{2}_d),
           \nonumber \\
    \textbf{8}_c
    &=&    (\textbf{2}_a \otimes \textbf{2}_d) \oplus (\textbf{2}_b \otimes \textbf{2}_c).
    \eea 
  Therefore, the S-duality permutes four $SU(2)$ symmetries.
  The strongly coupled limits of the original theory are S-dual to 
  weakly coupled limits of the theory where $SU(2)$ flavor symmetries are permuted.
  
  The generalization to the quiver gauge theory is straightforward.
  All possible superconformal quiver gauge theories with $SU(2)$ gauge groups can be constructed 
  from the fundamental building block $\CT_{0, 3}$ by gauging some of the flavor symmetries.
  The gauging of one $SU(2)$ symmetry of $\CT_{0, 3}$
  leads to two fundamental hypermultiplets, as seen above.
  By gauging two $SU(2)$ symmetries of $\CT_{0, 3}$, 
  we obtain a bifundamental hypermultiplet which has one $SU(2)$ flavor symmetry.
  Let us denote a bifundamental hypermultiplet by $B^{\alpha_1 \alpha_2 i}$, 
  where $i$ ($=1, 2$) are the flavor indices and $\alpha_1$ and $\alpha_2$ label two gauged $SU(2)$ respectively.
  The superpotential of this bifundamental can be written as
    \bea
    W
     =     B^{\alpha_1 \alpha_2 i} \left[ (\varepsilon \phi_1)_{\alpha_1 \beta_1} \varepsilon_{\alpha_2 \beta_2}
         + \varepsilon_{\alpha_1 \beta_1} (\varepsilon \phi_2)_{\alpha_2 \beta_2} \right] B^{\beta_1 \beta_2 j} 
           \varepsilon_{ij}.
    \eea
  Also, by gauging three $SU(2)$ flavor symmetries, we obtain a trifundamental multiplet.
  If we denote this multiplet by $T^{\alpha_1 \alpha_2 \alpha_3}$, the superpotential is
    \bea
    W
     =     T^{\alpha_1 \alpha_2 \alpha_3} 
           \left[ (\varepsilon \phi_1)_{\alpha_1 \beta_1} \varepsilon_{\alpha_2 \beta_2} \varepsilon_{\alpha_3 \beta_3}
         + \varepsilon_{\alpha_1 \beta_1} (\varepsilon \phi_2)_{\alpha_2 \beta_2} \varepsilon_{\alpha_3 \beta_3} 
         + \varepsilon_{\alpha_1 \beta_1} \varepsilon_{\alpha_2 \beta_2} (\varepsilon \phi_3)_{\alpha_3 \beta_3} \right] 
           T^{\beta_1 \beta_2 \beta_3}.
    \eea
  In order to obtain a superconformal gauge theory where $\beta$-functions of the gauge couplings vanish, 
  each gauge factor has to couple effectively to four fundamental hypermultiplets.
  This means that each $SU(2)$ gauge group should be obtained 
  by gauging of a diagonal subgroup of two $SU(2)$ flavor symmetries of $\CT_{0, 3}$'s.
  Collecting these pieces, we can construct various $\CN=2$ superconformal quiver gauge theories
  which have $SU(2)^p$ gauge group and $SU(2)^n$ flavor symmetry.
  Corresponding quiver diagrams have genus $g = 1 + \frac{p - n}{3}$.
  Applying the S-duality of $\CT_{0, 4}$ to each gauge group,
  we have different quiver gauge theories in $\CT_{g, n}$.
  (This procedure corresponds to $s$-$t$ duality 
  regarding the $SU(2)$ gauge factor as a propagator and $\CT_{0, 3}$ as a vertex.)
  Therefore, all the theories in $\CT_{g, n}$ are related by the S-dualities.
  
\begin{figure}[t]
    \begin{center}
    \includegraphics[scale=0.6]{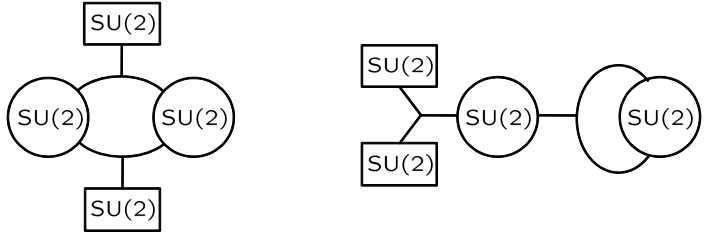}
    \caption{{\small Two different quivers $\CT_{1,2}$ which are related by the S-duality.}}
    \label{fig:KW}
    \end{center}
    \end{figure}

  In general, the global symmetry of $\CT_{g, n}$ is $SU(2)^n \times SU(2)_R \times U(1)_R$.
  However, in some cases, the flavor symmetry of the gauge theory is further enhanced.
  A trivial case is, of course, $\CT_{0, 4}$ where $SU(2)^4$ is enhanced to $SO(8)$.
  The simplest non-trivial case is $\CT_{1, 2}$, as depicted in Fig.~\ref{fig:KW}.
  The left quiver can be regarded as an $SO(4)$ gauge theory with four half-hypermultiplets transforming in $\textbf{4}$.
  Since $\textbf{4}$ is real, the flavor symmetry is enhanced to $USp(4)$.
  More concrete observation of the enhancement of the flavor symmetry, 
  based on $SU(2) \times SU(2)$ instead of $SO(4)$, will be presented in appendix \ref{sec:USp4}.
  
\begin{figure}[t]
    \begin{center}
    \includegraphics[scale=0.6]{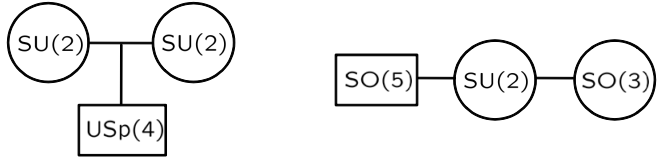}
    \caption{{\small The enhanced flavor symmetry of $\CT_{1, 2}$.
             Both quivers have the $USp(4) \cong SO(5)$ flavor symmetry.}}
    \label{fig:KWsymmetry}
    \end{center}
    \end{figure}

  The nontrivial S-dual theory of the above-mentioned one corresponds to the right quiver in Fig.~\ref{fig:KW}.
  The full flavor symmetry can be seen as follows:
  the right trifundamental is charged under the right $SU(2)$ gauge symmetry as 
  $\textbf{2} \otimes \textbf{2} = \textbf{3} \oplus \textbf{1}$.
  We can regard this as a $SO(3)$ vector and a singlet.
  Therefore, the trifundamental decomposes into a bifundamental of the $SU(2) \times SO(3)$ gauge symmetry
  and a fundamental of $SU(2)$.
  The latter is mixed with the left fundamentals to form an $SO(5)$ flavor symmetry (Fig.~\ref{fig:KWsymmetry}).
  This matches with the symmetry of the original quiver.
  We will analyze these in the subsequent sections more explicitly.
    
    
\begin{figure}[t]
    \begin{center}
    \includegraphics[scale=0.6]{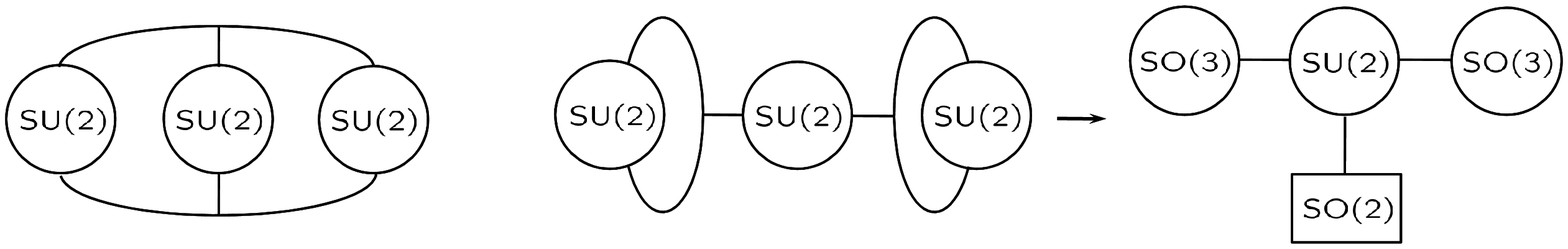}
    \caption{{\small The enhanced flavor symmetry of $\CT_{2, 0}$.
             Both quivers have the $SO(2)$ flavor symmetry.
             On the left, upper and lower trifundamentals form a $SO(2)$ vector.
             In the middle, each trifundamental decomposes as 
             $\textbf{2} \otimes \textbf{2} = \textbf{3} \oplus \textbf{1}$, as the $\CT_{1, 2}$ case.
             It produces two fundamental chiral multiplets for the center node (right).
             Thus, the flavor symmetry is $SO(2)$.}}
    \label{fig:T02}
    \end{center}
    \end{figure}

  Another example is $\CT_{2, 0}$.
  In this case, we can see that the flavor symmetry is enhanced to $SO(2)$ as in Fig.~\ref{fig:T02}.

  Finally, we comment on punctured Riemann surfaces which play a role in the study the $\CN =2$ quiver gauge theories.
  In \cite{Gaiotto}, it was pointed out that the space of the gauge coupling constants of the theory is identified 
  with the complex moduli space of the associated Riemann surface.
  The labels $g$ and $n$ of this family of SCFTs are precisely 
  the genus and the number of punctures of the corresponding Riemann surface. 
  Let us consider the schematic description of $\CT_{0,4}$ for instance, 
  which is associated with a sphere with four punctures $C_{0,4}$.
  The four $SU(2)$ flavor symmetries of $\CT_{0,4}$ correspond to four punctures on the sphere $C_{0,4}$.
  By the decoupling of the gauge coupling $\tau \rightarrow i \infty$, 
  we obtain two copies of $\CT_{0, 3}$'s, each of which has an $SU(2)^3$ flavor symmetry.
  This decoupling limit corresponds to the degeneration limit of a sphere into two three punctured-spheres.
  Thus, $\CT_{0,3}$ is the building block for the quiver, associated with a sphere with three punctures $C_{0,3}$.
  A decomposition of a punctured surface $C_{g,n}$ into $C_{0,3}$'s corresponds 
  to a weakly-coupled gauge theory description of the SCFTs $\CT_{g,n}$ as well.
  
    \begin{figure}[t]
    \begin{center}
    \includegraphics[scale=0.6]{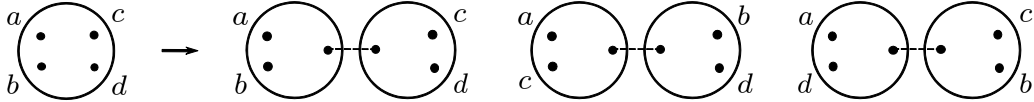}
    \caption{{\small Possible degeneration limits of a sphere with four punctures, which correspond to
             the usual weak coupling limit and S-dual weak coupling descriptions.}}
    \label{fig:T40decoupling}
    \end{center}
    \end{figure}    
    
 We can also give a schematic explanation of the S-dualities.
 Let us introduce the mass parameters for $\CT_{0,4}$ theory associated with $SU(2)_{a, b, \ldots}$ flavor symmetries.
 In this case, the gauge coupling moduli space is now described by a sphere with four marked punctures.
 Since the punctures are marked, we have three different degeneration limits of a sphere, 
 as in Fig.~\ref{fig:T40decoupling}.
 This is precisely the S-duality which permutes four $SU(2)$ symmetries.
 In generic $\CT_{g, n}$, all the possible weak coupling S-dual descriptions correspond 
 to the possible degeneration limits of the Riemann surface $C_{g, n}$.
    
\subsection{$\CN=1$ superconformal quiver gauge theories}
\label{subsec:SCFT}
  In what follows, we will consider the deformation of $\CN=2$ superconformal quiver gauge theories $\CT_{g, n}$ 
  to $\CN=1$ by the mass terms for the adjoint $\CN=1$ chiral multiplets in $\CN=2$ vector multiplets.
  
  Let us consider the $\CT_{0, 4}$ case, i.e., the $SU(2)$ gauge theory with four fundamental hypermultiplets.
  (For simplicity, we focus on the massless fundamental case.)
  With the mass deformation, below the energy scale of the mass parameter
  we integrate out the adjoint and the superpotential becomes $W \sim h Q^4$.
  In the ultraviolet, $h$ is an irrelevant coupling.
  However, it is, in the infrared, exactly marginal coupling in this $\CN=1$ theory \cite{LS}.
  Indeed, the $\beta$-function for $h$ is proportional to the one for the gauge coupling constant:
    \bea
    \beta_h \propto \beta_g \propto 1 + 2 \gamma,
    \eea
  where $\gamma$ is the anomalous dimension of $Q$.
  Thus, the solutions to $\beta_g = \beta_h = 0$ form one complex dimensional
  manifold (fixed line) because $\gamma$ is a function of $g$ and $h$.

  The existence of the fixed line can be convinced by considering $h = 0$, 
  that is $\CN=1$ SQCD with $N_f = 2 N_c$, in which it was shown that a nontrivial infrared fixed point exists \cite{Seiberg}.
  From $\CN=2$ point of view, the gauge coupling constant in the ultraviolet region, 
  which is the exactly marginal coupling in $\CT_{0, 4}$ theory, parametrizes this $\CN=1$ fixed line.
  
  The $\CT_{0, 4}$ theory is self-dual in a sense that 
  the S-duality does not change the quiver diagram. 
  Thus, we obtain a similar theory by the mass deformation \cite{LS}.
  This ``dual'' theory also has a quartic coupling $h_D$, whose value at the infrared fixed point
  is roughly the inverse of that of the original coupling $h$, 
  associated with the S-dual transformation of the gauge coupling constant.
  
  Things become more interesting when we consider higher $g$ and $n$ cases.
  As the $\CT_{0, 4}$ case above, $\CN=1$ quiver gauge theories obtained from $\CT_{g, n}$
  by the mass deformation are expected to flow to the superconformal fixed point.
  Furthermore, it leads to the higher dimensional manifold of fixed points, 
  which is associated with the fixed manifold of $\CN=2$ superconformal theories $\CT_{g, n}$.
  Partial evidence of the superconformal fixed points can be seen, as above, 
  by taking $h_{1, 2, \ldots, n} = 0$ and $g_{1, 2, \ldots, n} = 0$ except for $g_i$.
  In this case, the theory reduces to $\CN=1$ SQCD and we know the existence of the infrared fixed point.
  
  Recall that the S-dualities relate many different looking quiver gauge theories in $\CT_{g, n}$.
  Hence, we obtain many different looking $\CN=1$ quiver gauge theories by the mass deformation.
  These are supposed to describe the same infrared physics. 
  Therefore, this implies the existence of dualities among those $\CN=1$  quiver gauge theories.
  Of course, it includes the self-dual duality as the $\CT_{0, 4}$ case as well.
  
  The easiest non-trivial check of the existence of the superconformal fixed
  points and this Seiberg duality might be the 't Hooft anomaly matching.
  For generic quivers obtained from $\CT_{g, n}$ where the enhancement of the flavor symmetry does not occur, 
  the anomaly matching is very simple.
  We perform this at the origin of the moduli space of vacua.
  The global symmetry is $SU(2)_1 \times SU(2)_2 \times \ldots \times SU(2)_n \times U(1)_R$.
  The $SU(2)_i^3$ anomalies ($i = 1, \ldots, n$) vanish, 
  because the fundamental representation of $SU(2)$ are pseudo-real.
  Also, the $SU(2)_i \times U(1)_R^2$ anomalies are trivially zero.
  Next note that all the chiral multiplets have the same $U(1)_R$ charge $- 1/2$
  and there exist the same number of such chiral superfields for fixed $n$ and $g$ 
  if we do not distinguish the $SU(2)$ flavor and gauge indices.
  Note also that the multiplets which have the $SU(2)_i$ flavor symmetry, which are $\CT_{0, 3}$, 
  have the same number of extra indices of $SU(2)^2$.
  Therefore, $SU(2)_i^2 \times U(1)_R$, $U(1)_R^3$ and $U(1)_R$ anomalies are, respectively, 
  the same for fixed $n$ and $g$.
  
  The non-trivial case is the quivers obtained from $\CT_{1, 2}$ where the global symmetries are enhanced
  to $USp(4) (\cong SO(5)) \times U(1)_R$.
  The left quiver of Fig.~\ref{fig:KWsymmetry} has the chiral multiplets in $\textbf{4}$ of $USp(4)$,
  while the right quiver of Fig.~\ref{fig:KWsymmetry} has those in $\textbf{5}$ of $SO(5)$.
  Since these representations are real, the $USp(4)^3$ anomalies are zero for both sides.
  Also, since all the chiral multiplets have the same $U(1)_R$ charge $- 1/2$ as noted above, 
  the matching of the $U(1)_R^3$ and $U(1)_R$ anomalies is trivial.
  Finally, note that in terms of $USp(4)$, the quadratic Casimir of $\textbf{5}$ is twice as that of $\textbf{4}$.
  Therefore, 
    \bea
    (USp(4))^2 U(1)_R:
    ~~~~1 \times 4 \times \left( - \frac{1}{2} \right)
     =     2 \times 2 \times \left( - \frac{1}{2} \right).
    \eea
  This confirms that the anomalies of both theories match.

  In the rest of this paper, we will devote to collect other
  non-trivial evidence of the existence of the superconformal fixed
  points and the Seiberg dualities.
  In section \ref{sec:exactlymarginal}, we will consider exactly marginal operators 
  in these $\CN=1$ quiver gauge theories.
  In section \ref{sec:KW}, we concentrate on a particular example: 
  the generalized Klebanov-Witten theory and its dual introduced above.
  We will see the matching of the chiral operators and the nonlinear constraints on them.

\section{Exactly marginal operators}
\label{sec:exactlymarginal}
  In this section, we analyze exactly marginal operators 
  in $\CN=1$ quiver gauge theories with generic quartic terms in the superpotential.
  First of all, we briefly review the argument of \cite{LS} for the existence of the exactly marginal operators.
  Let us consider a supersymmetric gauge theory with product gauge groups, $\prod_{i=1}^{p} G_i$,
  and chiral multiplets, $\phi_a$, which is transformed as a representation $R_a(G_i)$ of the gauge group $G_i$.
  We consider a superpotential $W = \sum_s h_s W^{(s)}(\phi_a)$,
  where each $W^{(s)}$ is a product of $d_s$ chiral superfields ($s = 1, \ldots, m$).
  As in \cite{LS}, at a superconformal fixed point, the scaling coefficients 
    \bea
    A_{g_i}
    &=&  - \left( 3 C_2(G_i) - \sum_a T(R_a(G_i)) (1 - \gamma_a) \right),
           \\
    A_{h_s}
    &=&    d_s - 3 + \frac{1}{2} \sum_a \gamma_a \frac{\partial \ln W^{(s)}}{\partial \ln \phi_a}
           \label{beta2}
    \eea
  have to vanish \cite{LS}. 
  Here $C_2(G_i)$ is the quadratic Casimir, $T(R_a(G_i))$ is the index of the
  representation $R_a(G_i)$ and $\gamma_a$ is the anomalous dimension of the chiral superfield $\phi_a$.
  The derivative of the last term in (\ref{beta2}) counts the number of $\phi_a$ in $W^{(s)}$.
  From these, in general, we obtain $p + m$ equations.
  However, some of the equations would be degenerate.
  Let us denote the number of the linearly independent equations as $q$ ($\leq p + m$).
  These impose the $q$ conditions on $p + m$ coupling constants.\footnote{
    Of course there may be relevant and irrelevant operators which make the equations insolvable. 
    In this case, we take these coupling vanish and reconsider the equations 
    for the vanishing beta functions forgetting these couplings.}
  Thus, we expect that there is $p + m - q$ dimensional space of the solutions to these equations.
  In other words, there will exist $p + m - q$ exactly marginal operators.
  We will simply assume there indeed exist $p + m - q$ exactly marginal operators in our application.
  
  Now, we consider $\CN=1$ supersymmetric quiver gauge theories 
  associated with $\CN=2$ superconformal quiver gauge theories $\CT_{g, n}$.
  In particular, we consider the theories where the enhancement of the flavor symmetry does not occur.
  (We will analyze the theories with enhanced flavor symmetries, after the general discussion.)
  These theories are obtained as follows.
  First of all, we focus on a particular node of quiver.
  For such a node, the superpotential is
    \bea
    W
     =     \frac{1}{2} m \Tr \phi^2 + \sum_{s = 1,2} h_s \Tr \phi X_s,
    \eea
  where $\phi$ is the adjoint chiral superfield of the node we are considering.
  The second term is due to the superpotential of the $\CN=2$ supersymmetric gauge theory.
  The couplings $h_s$ are related with the gauge coupling $g$ in the $\CN=2$ theory.
  Each $X_s$ is $Q^2$ or $B^2$ or $T^2$, where $Q$, $B$, and $T$ are
  fundamental, bifundamental, and trifundamental superfields, respectively, 
  as seen in the subsection \ref{subsec:Gaiotto}.
  The trace is taken over the gauge indices of the node.
  All the other indices of the gauge and flavor symmetries of (bi or tri)fundamentals
  are already contracted as the superpotential is invariant under such symmetries.
  Integrating $\phi$ out, we obtain
    \bea
    W
    &=&  - \frac{1}{2 m} \left( h_1 \Tr X_1^2 + h_2 \Tr X_2^2 + 2 h_1 h_2 \Tr X_1 X_2 \right).
           \label{W1}
    \eea
  For each node, we add the mass term for the adjoint chiral field 
  and we have the superpotential (\ref{W1}) after integrating out it.
  The resulting $\CN=1$ supersymmetric gauge theory will be superconformal 
  at least if we tune the masses and the gauge coupling constants.
  Actually, all beta functions vanish if $\gamma_a=-1/2$ and there will be exactly marginal operators.
  
  We then consider how many exactly marginal operators keeping the flavor symmetry $SU(2)^n$ exist for this theory. 
  Because $\gamma_a=-1/2$, only quartic superpotential can be marginal.
  As we saw in subsection \ref{subsec:Gaiotto}, each node couples to two different matter multiplets, 
  say $P$ and $R$, each of which is two fundamentals or a bifundamental or a trifundamental.
  Let $X_1$ and $X_2$ be products of $P$ and $R$ respectively.
  The generic quartic superpotential keeping the flavor symmetry is a sum of the superpotential 
  associated for each node, like (\ref{W1}),
    \bea
    W
    &=&    H_1 P^4 + H_2 R^4 + H_3 P^2 R^2.
           \label{W}
    \eea
  Note that there is one independent quartic operator which is constructed from one field, 
  e.g. $P^4$ or $R^4$ in (\ref{W})\footnote{
    This fact is easily seen by noticing that the chiral field have three global or gauge $SU(2)$ indices, 
    let us denote it $Q^{a_1,a_2,a_3}$, 
    and there is only one invariant under the three $SU(2)$ constructed from four $Q$'s.}
  and in the sum of the superpotentials for the nodes,
  a $H_1$ or $H_2$-type coupling can appear in the superpotentials for two or three adjacent nodes.
  However, we need a bit care about the operator which is generated by two fields, like $P^2 R^2$ in (\ref{W}).
  Generically, there exists one such independent operator from two fields
  and, therefore, the $H_3$ type coupling appear only once in the superpotential at a node.
  In special cases, however, it is possible to construct two independent operators.
  For a moment, we assume that the number of such independent operator is one
  and treat such cases after general argument.

  In order to count the exactly marginal operators for the $\CN=1$ theory,
  it is convenient to consider the relations between scaling coefficients when the mass perturbation is turned off.
  Let $A_{\tilde{g}}$ and $A_{h_s}$ be the scaling coefficients of the gauge coupling constant
  and the couplings $h_s$ before integrating out $\phi$.
  We refer to the anomalous dimensions of $P$, $R$ and $\phi$ as $\gamma_P$, $\gamma_R$ and $\gamma_\phi$, respectively.
  In this case, the scaling coefficients for $h_1$ and $h_2$ are 
  $A_{h_1} = \frac{1}{2} (\gamma_\phi + 2 \gamma_P)$ and
  $A_{h_2} = \frac{1}{2} (\gamma_\phi + 2 \gamma_R)$.
  In terms of these, $A_{\tilde{g}}$ can be written as
    \bea
    A_{\tilde{g}}
     =   - 2 (\gamma_\phi + \gamma_P + \gamma_R)
     =   - 2 (A_{h_1} + A_{h_2}).
           \label{Ag2}
    \eea
  Such relation is satisfied for each node.
  
  Then, we return to consider the $\CN=1$ supersymmetric quiver gauge theory with (\ref{W}).
  The scaling coefficients for $H_1$, $H_2$ and $H_3$ can be evaluated as
    \bea
    A_{H_1}
    &=&    1 + \left( A_{h_1} - \frac{1}{2} \gamma_\phi \right)
         + \left( A_{h_1} - \frac{1}{2} \gamma_\phi \right)
     =     1 + 2 A_{h_1} - \gamma_\phi,
           \nonumber \\
    A_{H_2}
    &=&    1 + 2 A_{h_2} - \gamma_\phi,
           \nonumber \\
    A_{H_3}
    &=&    1 + A_{h_1} + A_{h_2} - \gamma_\phi
     =     \frac{1}{2} (A_{H_1} + A_{H_2}).
           \label{AH3}
    \eea
  Also, by using (\ref{Ag2}), the scaling coefficient for the gauge coupling $g$ for the node can be calculated as
    \bea
    A_{g}
     =     A_{\tilde{g}} - 2 (1 - \gamma_\phi)
     =   - 2 (A_{h_1} + A_{h_2} + 1 - \gamma_\phi)
     =   - (A_{H_1} + A_{H_2}).
           \label{Ag1}
    \eea
  Therefore, we see that two equations, (\ref{AH3}) and (\ref{Ag1}), are redundant.
  
  So far, we only considered a particular node.
  For each node, we can do the same calculation as above.
  As we assumed above, 
  the operator associated with the $H_3$-type coupling constant appears exactly one time for each node.
  Therefore, the number of the dependent equations is 2 for each node
  and we conclude that the number of the exactly marginal operator is $2p$, 
  where $p$ is the total number of the nodes.
  
    \begin{figure}[t]
    \begin{center}
    \includegraphics[scale=0.6]{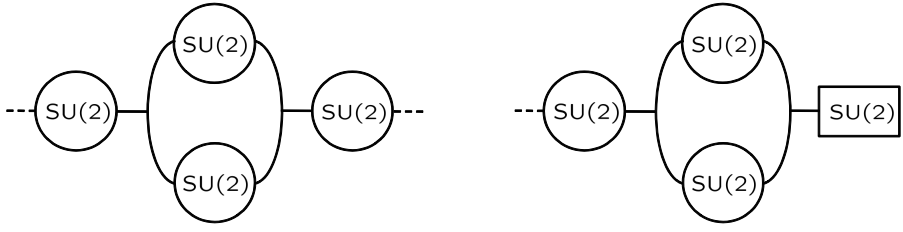}
    \caption{{\small A quiver including ``small loop" which consists of two gauge groups and two tri(or bi)fundamentals.}}
    \label{fig:smallloop}
    \end{center}
    \end{figure}
  
  At this stage, let us analyze the validity of the assumption.
  We should be careful about the counting of independent $H_3$-type operators
  in the case where the quiver includes ``small loop", 
  which consists of two nodes and two bi(or tri)fundamentals as depicted in Fig.~\ref{fig:smallloop}.
  One might think that the $H_3$-type operators 
  from the upper node and that from the lower node are the same.
  However, we can construct two independent $H_3$-type operators in this case.
  Therefore, there are two $H_3$-type operators for two nodes of the small loop
  and the conclusion in the previous paragraph is correct.
  
\begin{figure}[t]
    \begin{center}
    \includegraphics[scale=0.6]{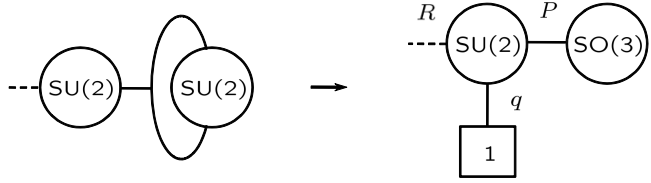}
    \caption{{\small The quiver including a loop at the tip of quiver (left).
             As in Fig.~\ref{fig:KWsymmetry}, this quiver can be seen as the right hand side, 
             where $1$ means one chiral multiplet.}}
    \label{fig:legsmallloop}
    \end{center}
    \end{figure}

  There is one more exceptional case which needs care. 
  This is the quivers with the leg ending by the small loop, as Fig.~\ref{fig:legsmallloop}.
  Naive consideration leads to that the $H_3$-type operator for the right hand side node is
  the same as the $H_1$- or $H_2$-operator.
  It follows that the number of the exactly marginal operators is reduced by one for such a loop.
  However, this naive conclusion is wrong.
  Indeed, we can see this quiver as the right hand side quiver in Fig.~\ref{fig:legsmallloop}.
  In that, we have three different fields, a bifundamental chiral superfield of $SU(2) \times SO(3)$, $P$, 
  a fundamental chiral superfield for $SU(2)$, $q$, and a chiral superfield, $R$.
  Let $\gamma_P$, $\gamma_q$ and $\gamma_R$ be the anomalous dimensions of $P$, $q$ and $R$.
  The scaling coefficients for the gauge couplings are
    \bea
    A_{g_1}
     =   - \left( 2 + 2 \gamma_R + \frac{3}{2} \gamma_P + \frac{1}{2} \gamma_q \right),
           ~~~
    A_{g_2}
     =   - 2 (1 + 2 \gamma_P),
    \eea
  where $g_1$ and $g_2$ are the gauge couplings of the $SU(2)$ and $SO(3)$ groups.
  Also, the generic quartic superpotential from these fields is schematically
    \bea
    W
     =     \lambda_1 P^4 + \lambda_2 R^4 + \lambda_3 R^2 P^2 + \lambda_4 q^2 R^2 + \lambda_5 q^2 P^2.
    \eea
  The scaling coefficients are
    \bea
    A_{\lambda_1}
    &=&    1 + 2 \gamma_P, 
           ~~~
    A_{\lambda_2}
     =     1 + 2 \gamma_R,
           ~~~
    A_{\lambda_3}
     =     1 + \gamma_P + \gamma_R,
           \nonumber \\
    A_{\lambda_4}
    &=&    1 + \gamma_q + \gamma_R,
           ~~~
    A_{\lambda_5}
     =     1 + \gamma_q + \gamma_P.
    \eea
  Among them, four constraints are redundant.
  Therefore, there are four exactly marginal operators and it matches with the general rule above. 
  We conclude 
  that the number of the exactly marginal operators are same for a class of theories which are Seiberg dual each other.
  
  We note that the Seiberg dual theory contains the gauge singlet, meson, for $\CN=1$ SQCD
  and expect that there are theories with mesons for our cases.
  In the discussion of \cite{LS}, the theory containing the singlet meson was introduced and 
  was expected to flow to the infrared superconformal fixed point of $\CN=1$ SQCD with the quartic superpotential.
  In our case, the quartic superpotential for each node is
    \bea
    W
     =     \frac{H_1}{2} P^4 + \frac{H_2}{2} R^4 + \frac{H_3}{2} P^2 R^2.
           \label{superpotentialH}
    \eea
  ($H_1$ and $H_2$ type operators will appear in the superpotentials from different nodes.)
  We can also introduce the theory with mesons which has the same chiral operators as follows:
    \bea
    W 
     =     N_1 P^2 + N_2 R^2 + N_3 P R - \frac{1}{2H_1} N_1^2 - \frac{1}{2 H_2} N_2^2 - \frac{1}{2 H_3} N_3^2,
    \eea
  where $N_{1, 2, 3}$ are independent operators 
  and, in $P^2$, $R^2$ and $PR$, only the gauge indices of the node we are considering are contracted.
  We expect them to flow to the infrared superconformal fixed point of the theory with no meson (\ref{superpotentialH}),
  although there is no strong evidence to support it.
  At least the chiral ring of the theory with mesons is the same as the one without mesons.
  
  We have counted the number of the exactly marginal operators keeping the flavor symmetry $SU(2)^n$.
  However, as seen in subsection \ref{subsec:Gaiotto}, the flavor symmetry would be enhanced in some cases.
  The number of the exactly marginal operator keeping this enhanced symmetry could be reduced. 
  Let us consider these cases here.
  The first simple example is the $SU(2)$ gauge theory with four flavors (eight fundamental chiral multiplets) 
  obtained from $\CT_{0, 4}$, where the flavor symmetry is $SO(8)$.
  As noted in subsection \ref{subsec:SCFT}, 
  the coupling constant of a quartic term in the superpotential is a exactly marginal coupling
  and there is one exactly marginal operator.
  Indeed, the independent quartic operator is unique in this case and 
  two conditions $\beta_g = \beta_h = 0$ are linearly dependent.
  
  A nontrivial case is the quivers obtained from $\CT_{1, 2}$ in Fig.~\ref{fig:KWsymmetry}.
  As analyzed in subsection \ref{subsec:Gaiotto}, these two quivers have the $USp(4) \cong SO(5)$ flavor symmetry.
  First of all, let us consider the quiver associated with the left hand side in Fig.~\ref{fig:KWsymmetry}, 
  which we will call as ``generalized" Klebanov-Witten theory in the following section.
  (The meaning of ``generalized" will be soon clear.)
  As we will analyze in section \ref{sec:KW}, there are two independent quartic operators:
  a bilinear of the mesonic operator and a baryonic one.
  Therefore, the most generic superpotential is
    \bea
    W
     =     h_1 J_{ik} J_{jl} Q^{\mu i} Q^{\mu j} Q^{\nu k} Q^{\nu l}
         + h_2 \varepsilon_{\mu \nu \rho \sigma} \varepsilon_{ijkl} Q^{\mu i} Q^{\nu j} Q^{\rho k} Q^{\sigma l}.
           \label{superpotentialT12}
    \eea
  In this case, all the scaling coefficients are proportional to $1 + 2\gamma_Q$,
  where $Q$ is the anomalous dimension of $Q$.
  Therefore, there are three exactly marginal operators keeping $USp(4)$ in this theory.\footnote{
    The reader may wonder that this counting is different from that in Klebanov-Witten theory \cite{KW}. 
    This is because Klebanov-Witten theory was obtained by a specific mass deformation from $\CT_{1,2}$, 
    where mass parameters are chosen as $m_1 = - m_2$, as we will see explicitly in section \ref{sec:KW}.
    In that case, the first term in (\ref{superpotentialT12}) vanishes and the flavor symmetry is enhanced to $SU(4)$.
    Hence, the number of the exactly marginal operators is reduced to two.
    This matches with the result in \cite{KW}. 
    This enhancement of the flavor symmetry should be seen in the dual theory non-trivially 
    although we have not find a mechanism yet.}
  On the other hand, the right hand side quiver in Fig.~\ref{fig:KWsymmetry} has different matter content.
  We will refer to this theory as dual theory in the following.
  This quiver is very similar to the one in Fig.~\ref{fig:legsmallloop}.
  Therefore, the counting of the exactly marginal operators is straightforward
  and we obtain the same answer as the generalized Klebanov-Witten theory.
  
  We also consider two quivers obtained from the $\CT_{2, 0}$ theories (Fig.~\ref{fig:T02}),
  where the flavor symmetry is enhanced to $SO(2)$.
  There are six exactly marginal operators in both two theories.
  (Note that if we keep the $SU(2)^n$ flavor symmetry, that is no flavor symmetry, 
  the number of the exactly marginal operators is nine.
  This is only case where the general rule for the number is incorrect.)
  
  Finally, we comment on the $SU(N)^p$ gauge group case. 
  In the case where the Lagrangian descriptions are exist, e.g., $A_p$ and $\hat{A}_{p-1}$ theories 
  \cite{Cachazo:2001gh, Cachazo:2001sg, DM}, we can follow above argument.
  There are many exactly marginal operators.\footnote{
    We thank the referee for pointing out the mistakes in the previous version.}
  We expect that the theories flow to non-trivial infrared fixed points.

\section{Generalized Klebanov-Witten theory and its dual}
\label{sec:KW}
  In this section, we consider a simple example which leads to a non-trivial Seiberg duality by the mass deformation.
  Since a theory with one gauge group does not have a non-trivial S-duality which changes matter contents,
  we consider a theory with two gauge groups $SU(2) \times SU(2)$.  
  In particular, we consider a theory with genus one, 
  which is specified by the generalized quivers illustrated in Fig.~\ref{fig:KW}.
  The claim is that when we consider the mass deformation for each theory, the resulting theories are Seiberg dual.
  
  In the following, we consider the mass deformation for each theory.
  From the quiver illustrated in the left of Fig.~\ref{fig:KW}, 
  we obtain the $\CN=1$ generalized Klebanov-Witten theory
  as discussed in the previous section.
  The other quiver (the right in Fig.~\ref{fig:KW}) leads to its dual theory.
  We will see that the chiral rings of the resulting theories match non-trivially, 
  e.g., a non-linear constraint for the mesons from the F-term equation is dual to a classical trivial constraint.
  Note that we analyze the classical chiral rings and classical moduli spaces. 
  Since for $|M| \gg \Lambda$, the gauge group is generically broken to an abelian group for these theories, 
  no non-perturbative effect will appear 
  unlike the mass deformed $\CN=1$ SQCD with $N_f = N_c + 2$ (in the dual description) \cite{Seiberg}.
  Therefore, this matching can be regarded 
  as non-trivial evidence of the existence of the superconformal field theory and the Seiberg dual.
  In subsection \ref{subsec:generalizedKW} and \ref{subsec:dual}, 
  we analyze the generalized Klebanov-Witten theory and its dual respectively.

\subsection{Generalized Klebanov-Witten theory}
\label{subsec:generalizedKW}
  In this subsection, we analyze the classical chiral ring of the generalized Klebanov-Witten theory,
  whose gauge group is $SU(2)_1 \times SU(2)_2$ and the flavor symmetry is $USp(4)$.
  The superpotential of the ${\cal N}=2$ theory before the mass deformation is given by
    \begin{align}
    W
     =     {\rm Tr} (B_1 \phi A_1) + {\rm Tr} (B_2 \phi A_2)
         + {\rm Tr} (A_1 \tilde{\phi} B_1) + {\rm Tr} (A_2 \tilde{\phi} B_2),
    \end{align}
  where $A$ and $B$ are bifundamental chiral superfields and
  $\phi$ ($\tilde{\phi}$) is an adjoint chiral superfields of $SU(2)_1$ ($SU(2)_2$).
  By adding the following mass term
    \begin{align}
    W
     =     \frac{1}{2} m_1 {\rm Tr} \phi^2 + \frac{1}{2} m_2 {\rm Tr} \tilde{\phi}^2, 
    \end{align}
  and integrating out the massive adjoint chiral superfields, we obtain
    \begin{align}
    W
     =     
    &      \frac{1}{8} \left( \frac{1}{m_1} + \frac{1}{m_2} \right)
           J_{ik} J_{jl} Q^{\mu i} Q^{\mu j} Q^{\nu k} Q^{\nu l} \nonumber \\
    &    - \frac{1}{2\cdot 4!} \left( \frac{1}{m_1} - \frac{1}{m_2} \right) 
           \varepsilon_{\mu \nu \rho \sigma} \varepsilon_{ijkl} Q^{\mu i} Q^{\nu j} Q^{\rho k} Q^{\sigma l}.
           \label{W_KW}
    \end{align}
  Here, for later convenience, we renamed the fields as 
    \begin{align}
    (A_1, A_2, B_1, B_2) \to (Q^1, Q^2, Q^3, Q^4),
    \end{align}
  and regard the gauge group as $SO(4)$ instead of $SU(2)_1 \times SU(2)_2$.
  We label the indices of the gauge group $SO(4)$ by $\mu,\nu, \cdots$ 
  and that of the global $USp(4)$ group by $i,j,\cdots$.
  The invariant tensor $J$ is defined as 
    \begin{align}
    J_{ij}
     =     \left( \begin{array}{cc}
           {\bf 0} & {\bf 1} \\
           {\bf -1} &  {\bf 0} \\
           \end{array}\right) , \qquad
    J^{ij}
     =     \left( \begin{array}{cc}
           {\bf 0} & {\bf -1} \\
           {\bf 1} &  {\bf 0} \\
           \end{array} \right),
    \end{align}
  where ${\bf -1},{\bf 0},{\bf 1}$ are $2\times 2$ matrices.
  For details about changing the notation from $SU(2)_1 \times SU(2)_2$ to $SO(4)$ 
  and about integrating out the adjoint superfields, see appendix \ref{sec:massdeformation}.
  Matter contents of this theory are summarized in Table \ref{matter_original}.

    \begin{table}
    \centering
    \begin{tabular}{c|c|c}
      & SO(4) & USp(4) \vspace{-3mm} \\
      &{\scriptsize $\mu, \nu, \cdots$} & {\scriptsize $i,j, \cdots$} \\
      \hline
      $Q^{\mu i}$ & {\bf 4}     &  {\bf 4}   \\ 
      $(W_{\alpha})^{\mu}{}_{\nu}$ & {\bf 6} & {\bf 1}
    \end{tabular}
    \caption{Matter contents of the generalized Klebanov-Witten theory.
     $W_{\alpha}$ is a field strength.}
    \label{matter_original}
    \end{table}

  It is known that for the $SO(4)$ gauge theory with the superpotential $W=0$, 
  the independent operators in the chiral ring are \cite{IS}
    \begin{align}
    &M^{(ij)} = Q^{\mu i} Q^{\mu j}, \\
    &B = \frac{1}{4!} 
    \varepsilon_{\mu \nu \rho \sigma} \varepsilon_{ijkl} 
    Q^{\mu i} Q^{\nu j} Q^{\rho k} Q^{\sigma l}, \\
    &h^{[ij]}_{\alpha} = \frac{1}{2} \varepsilon_{\mu \nu \rho \sigma} 
    \left( Q^{\mu i} Q^{\nu j} 
    - \frac{1}{4} (Q^{\mu k} Q^{\nu l} J_{lk}) J^{ij} \right)
    W_{\alpha}{}^{\rho \sigma}, \\
    &h_{\alpha} = \frac{1}{4} \varepsilon_{\mu \nu \rho \sigma} 
    J_{ij} Q^{\mu i} Q^{\nu j} W_{\alpha}{}^{\rho \sigma}, \\
    &H = \frac{1}{4} \varepsilon_{\mu \nu \rho \sigma} 
    W^{\alpha}{}^{\mu\nu} W_{\alpha}{}^{\rho\sigma}, \\
    &S= {\rm Tr} W^{\alpha} W_{\alpha}.
    \end{align}
  For later convenience, we have decomposed the operator $QQW$ into $h^{[ij]}_{\alpha}$ and $h_{\alpha}$, 
  which are in the irreducible representations of the global $USp(4)$ group.
  
  Since we actually have a non-vanishing superpotential, we have to consider the equations of motion:
    \begin{align}
    (m_1+m_2) J_{ik} J_{jl} Q^{\mu j} Q^{\nu k} Q^{\nu l}
    + \frac{1}{6}(m_1-m_2) \varepsilon_{\mu \nu \rho \sigma} \varepsilon_{ijkl} 
    Q^{\nu j} Q^{\rho k} Q^{\sigma l} = 0,
    \label{eom_original}
    \end{align}
  which leads to non-trivial chiral ring relations.
  By multiplying $Q^{\mu m} J^{ni}$ to the equations of motion (\ref{eom_original}), we obtain
    \begin{align}
    (m_1+m_2) M^{mj} J_{jl} M^{ln}
     =     (m_1-m_2) B J^{mn}.
    \end{align}
  This is decomposed into the following irreducible representations of $USp(4)$:
    \begin{align}
    & (m_1+m_2) \left[ M^{ij} J_{jk} M^{kl} - \frac{1}{4} \left( M^{mn} J_{np} M^{pq} J_{qm} \right) J^{il} \right]
     =     0 \label{con5}, \\
    & (m_1+m_2) M^{ij} J_{jk} M^{kl} J_{li} = 4 (m_1-m_2) B.
    \label{con1}
    \end{align}
  The first equations show non-linear constraints on the meson operator $M^{ij}$, 
  while the second equation indicates that the baryon operator $B$ is decomposed into the product of the meson operator.
  Only when $m_1-m_2=0$, the baryon operator exists in the chiral ring. 
  
  Other non-linear constraints can be obtained by multiplying 
  $\varepsilon_{\mu\kappa\lambda\tau} \varepsilon_{mnpq} Q^{\kappa n}Q^{\lambda p}Q^{\tau q}$ 
  to the equations of motion.
  Using identities among invariant tensors like
  $\varepsilon^{\mu\nu\rho\sigma} \varepsilon_{\mu\kappa\lambda\tau} 
  = \delta^{[\nu}_{\kappa} \delta^{\rho}_{\lambda} \delta^{\sigma]}_{\tau}$,
  we obtain a constraint
    \begin{align}
    (m_1+m_2) B J_{ij} M^{jk} J_{km} 
     =     \frac{1}{6} (m_1-m_2) ({\rm cof} M)_{im},
           \label{M3_original}
    \end{align}
  where cofactor ${\rm cof} M$ of the matrix $M$ is defined as
    \begin{align}
    ({\rm cof} M)_{im}
     \equiv 
           \varepsilon _{ijkl} \varepsilon_{mnpq} M^{jn} M^{kp} M^{lq}.
           \label{cofM}
    \end{align}
  Since the totally antisymmetric invariant tensor $\varepsilon_{ijkl}$ can be rewritten 
  in terms of the invariant tensor $J_{ij}$ as
    \begin{align}
    \varepsilon_{ijkl}
     =   - J_{i[j} J_{kl]},
    \end{align}
  (\ref{cofM}) can also be rewritten as 
    \begin{align}
    ({\rm cof} M)_{im}
     =     3 J_{ij} M^{jk} J_{km} (J_{np}M^{pq}J_{qr}M^{rn}) - 6 J_{ij} M^{jk} J_{kl} M^{ln} J_{np} M^{pq} J_{qm}.
    \end{align}
  By using this identity together with (\ref{con1}), the constraint (\ref{M3_original}) becomes
    \begin{align}
    & 4 (m_1 - m_2)^2 J_{ij} M^{jk} J_{kl} M^{ln} J_{np} M^{pq} J_{qm} \nonumber \\
    & -  \left[ 2(m_1-m_2)^2 - (m_1+m_2)^2 \right]
    J_{ij} M^{jk} J_{km} (J_{np}M^{pq}J_{qr}M^{rn})
     =     0.
           \label{JMJMJMJ}
    \end{align}
On the other hand, by multiplying $J_{ri} \times J_{ls} M^{st} J_{tu}$
to (\ref{con5}), we obtain
\begin{align}
(m_1+m_2)
\left[ J_{ri} M^{ij} J_{jk} M^{kl} J_{ls} M^{st} J_{tu}
- \frac{1}{4} \left( M^{mn} J_{np} M^{pq} J_{qm} \right) 
J_{rs} M^{st} J_{tu} \right] = 0.
\label{JMJMJMJ2}
\end{align}
Comparing (\ref{JMJMJMJ}) and (\ref{JMJMJMJ2}),
we obtain the following two constraints:
\begin{align}
J_{ij} M^{jk} J_{km} (J_{np}M^{pq}J_{qr}M^{rn}) = 0 \label{M1M2}, \\
J_{qi} M^{ij} J_{jk} M^{kl} J_{lm} M^{mn} J_{np} = 0 \label{M3},
\end{align}
  for generic masses. 
  Since we are assuming that $m_1\neq 0$, $m_2\neq 0$ in order that we can integrate out the adjoint field, 
  (\ref{JMJMJMJ}) and (\ref{JMJMJMJ2}) cannot be identical constraints.  
  Only the special case is $m_1+m_2=0$, where the constraints (\ref{JMJMJMJ2}) vanish and (\ref{JMJMJMJ}) becomes 
\begin{align}
J_{ri} M^{ij} J_{jk} M^{kl} J_{ls} M^{st} J_{tu}
- \frac{1}{2} \left( M^{mn} J_{np} M^{pq} J_{qm} \right) 
J_{rs} M^{st} J_{tu} = 0.
\end{align}

If we impose the former constraints (\ref{M1M2}), 
the latter (\ref{M3}) can be derived from
(\ref{JMJMJMJ2}), which originates in (\ref{con5}).
Thus, the independent constraints for generic masses are (\ref{con5}) and (\ref{M1M2}).

The usual classical constraints 
\begin{align}
B^2 = \frac{1}{4!} \det M
\end{align}
can be derived from (\ref{con1}) and (\ref{M3_original}),
and does not lead to a new constraint for the meson operator.

In summary, when the masses are generic,
the independent chiral operators are
$
M^{ij}, h^{[ij]}_{\alpha}, h_{\alpha}, H, S,
$
and the non-linear constraints for the meson operator are
\begin{align}
(MJM)^{ij} - \frac{1}{4} {\rm Tr} (MJMJ) J^{ij} = 0, 
\label{sum1} \\
{\rm Tr} (MJMJ) M^{ij} = 0. \label{sum2}
\end{align}
Other non-linear constraints for operators 
including $W_{\alpha}$ also exist,
but we do not analyze them here.

\subsection{Dual theory}
\label{subsec:dual}
  In this subsection, we consider the dual of the generalized Klebanov-Witten theory,
  whose generalized quiver is illustrated in Fig.~\ref{fig:KWsymmetry}.
  
  The superpotential of this dual ${\cal N}=2$ theory before the mass deformation is given by
    \begin{align}
    W
     =     P^a{}_b{}^{\dot{a}} \phi ^b{}_c P^c{}_a{}^{\dot{b}} \varepsilon_{\dot{a}\dot{b}}
         + P^a{}_b{}^{\dot{a}} \varepsilon_{\dot{a}\dot{b}} \tilde{\phi} ^{\dot{b}}{}_{\dot{c}} P^b{}_a{}^{\dot{c}} 
         + q^{\dot{a}}{}_I \varepsilon_{\dot{a}\dot{b}} \tilde{\phi}^{\dot{b}}{}_{\dot{c}} q^{\dot{c}}{}_I,
    \end{align}
  where $\phi$ is the adjoint field of $SU(2)_1$ and $\tilde{\phi}$ is the adjoint field of $SU(2)_2$.
  By adding the following mass term 
    \begin{align}
    W
     =     \frac{1}{2} m_1 {\rm Tr} \phi^2 + \frac{1}{2} m_2 {\rm Tr} \tilde{\phi}^2, 
    \end{align}
  and integrating out the massive adjoint fields, we obtain
    \begin{align}
    W
     =     &-\frac{1}{2} \left( \frac{1}{m_1} + \frac{1}{m_2} \right) 
           P^{a}{}_b{}^{\dot{a}} P^{b}{}_a{}^{\dot{b}} \varepsilon_{\dot{b} \dot{c}} 
           P^{c}{}_d{}^{\dot{c}} P^{d}{}_c{}^{\dot{d}} \varepsilon_{\dot{d} \dot{a}}
           \nonumber \\
    &    - \frac{1}{m_2} 
           q^{\dot{a}}_I \varepsilon_{\dot{a}\dot{b}} P^{a}{}_b{}^{\dot{b}} 
           P^{b}{}_a{}^{\dot{c}} \varepsilon_{\dot{c}\dot{d}} q^{\dot{d}}_I - \frac{1}{2m_2} 
           q^{\dot{a}}_I \varepsilon_{\dot{a}\dot{b}} q^{\dot{b}}_J 
           q^{\dot{c}}_J \varepsilon_{\dot{c}\dot{d}} q^{\dot{d}}_I.
    \end{align}
  Matter contents of this mass deformed dual theory is summarized in Table \ref{matter_dual}.

\begin{table}
\centering
\begin{tabular}{c|cc|c}
 & SU(2) & SU(2) & SO(5) \vspace{-3mm} \\
 & {\scriptsize $a,b,\cdots$} & {\scriptsize $\dot{a},\dot{b},\cdots$} 
 & {\scriptsize $I,J,\cdots$} \\
\hline
$P^{a}{}_b{}^{\dot{a}}$   & {\bf 3} & {\bf 2} & {\bf 1} \\ 
$q^{\dot{a}}{}_{I}$  & {\bf 1} & {\bf 2} & {\bf 5} \\
$W_{\alpha}{}^a{}_b$ & {\bf 3} & {\bf 1} & {\bf 1} \\ 
$w_{\alpha}{}^{\dot{a}}{}_{\dot{b}}$ 
& {\bf 1} & {\bf 3} & {\bf 1} \\ 
\end{tabular}
\caption{Matter contents of the dual theory.}
\label{matter_dual}
\end{table}

As discussed in appendix \ref{sec:chiraloperators}, the independent generators 
of the chiral ring of this theory for generic masses are
\begin{align}
&\tilde{M}_{IJ} = q^{\dot{a}}{}_{I} \label{M_IJ}
\varepsilon_{\dot{a} \dot{b}} q^{\dot{b}}{}_{J}, \\
& \tilde{h}^{I}_{\alpha} = q^{\dot{a}}_I \varepsilon_{\dot{a} \dot{b}} 
P^{a}{}_b{}^{\dot{b}} (W_{\alpha})^{a}{}_{b}, \\
& \tilde{h}_{\alpha} = P^{a}{}_b{}^{\dot{a}} \varepsilon_{\dot{a}\dot{b}}
P^{b}{}_c{}^{\dot{b}} (W_{\alpha})^c{}_a 
\,\, \sim \,\, P^{a}{}_b{}^{\dot{a}} P^{b}{}_a{}^{\dot{b}} 
\varepsilon_{\dot{b}\dot{c}} (w_{\alpha})^{\dot{c}}{}_{\dot{a}}, \\
& S_1 = {\rm Tr} \,\, W^{\alpha} W_{\alpha}, \\
& S_2 ={\rm Tr} \,\, w^{\alpha} w_{\alpha}, \label{S_2}
\end{align}
  where we used the equations of motion:
    \begin{align}
    &-2 \left( \frac{1}{m_1} + \frac{1}{m_2} \right) 
    P^{b}{}_a{}^{\dot{b}} \varepsilon_{\dot{b} \dot{c}} 
    P^{c}{}_d{}^{\dot{c}} P^{d}{}_c{}^{\dot{d}} \varepsilon_{\dot{d} \dot{a}}
    - \frac{2}{m_2} 
    q^{\dot{b}}_I \varepsilon_{\dot{b}\dot{a}} 
    P^{b}{}_a{}^{\dot{c}} \varepsilon_{\dot{c}\dot{d}} q^{\dot{d}}_I = 0, \label{_eom1_dual}\\
    &- \frac{2}{m_2} 
    \varepsilon_{\dot{a}\dot{b}} P^{a}{}_b{}^{\dot{b}} 
    P^{b}{}_a{}^{\dot{c}} \varepsilon_{\dot{c}\dot{d}} q^{\dot{d}}_I - \frac{2}{m_2} 
    \varepsilon_{\dot{a}\dot{b}} q^{\dot{b}}_J 
    q^{\dot{c}}_J \varepsilon_{\dot{c}\dot{d}} q^{\dot{d}}_I = 0. \label{_eom2_dual}
    \end{align}

These operators match to those of the original theory as 
\begin{align}
M^{ij} & \sim (\Gamma^{IJ})^i{}_k J^{kj} \tilde{M}_{IJ}, \\
h^{[ij]}_{\alpha} & \sim \tilde{h}^{I}_{\alpha} 
(\Gamma^{I})^i{}_k J^{kj}, \\
h_{\alpha} & \sim \tilde{h}_{\alpha}, \\
H, S &\sim S_1, S_2,
\end{align} 
where $\Gamma^{I}$ is the gamma matrices for the $Spin(5)$ group,
and chosen such that 
$$ (\Gamma^{I})^{(i}{}_k J^{|k|j)} = 0,
\qquad {\rm Tr} \Gamma^I = 0.$$
Certain linear combinations of $S_1$ and $S_2$ correspond to 
$H$ and $S$, which cannot be determined from the global charge. 
Only when $m_1+m_2=0$, an operator
\begin{align}
P^4 = P^{a}{}_b{}^{\dot{a}} P^{b}{}_a{}^{\dot{b}} \varepsilon_{\dot{b} \dot{c}} 
P^{c}{}_d{}^{\dot{c}} P^{d}{}_c{}^{\dot{d}} \varepsilon_{\dot{d} \dot{a}}
\end{align}
appears in the classical chiral ring.
This operator is expected to correspond to the baryon operator $B$ of 
the generalized Klebanov-Witten theory.
This matching of the operators in the classical chiral ring
gives a non-trivial consistency check to the duality.

We further investigate the non-linear constraints of the meson 
operator.
From the definition of the meson operator $\tilde{M}_{IJ}$, 
we obtain non-linear constraints
\begin{align}
\tilde{M}_{I[J} \tilde{M}_{KL]} = 0,
\label{MM}
\end{align}
which follow from the identity
$q^b{}_{[J} q^c{}_K q^d{}_{L]} =0 $,
where the gauge indices of $SU(2)_1$ run $a,b,c=1,2$.
This can also be rewritten as 
\begin{align}
\varepsilon^{IJKLM} \tilde{M}_{JK} \tilde{M}_{LM} = 0.
\label{MM0}
\end{align}
Further non-linear constraints can be obtained from 
equations of motion (\ref{_eom1_dual}) and (\ref{_eom2_dual}).
By multiplying $P^c{}_d{}^{\dot{c}}P^d{}_c{}^{\dot{d}}$ 
to the second equations of motion (\ref{_eom2_dual}),
and by using (\ref{P4_qPPq}) and (\ref{qPPq_MM}) in appendix \ref{sec:chiraloperators},
which are also derived from the equations of motion,
we finally obtain
\begin{align}
m_1 (\tilde{M}_{KL} \tilde{M}_{LK}) \tilde{M}_{IJ}
= 2(m_1+m_2) \tilde{M}_{IK} \tilde{M}_{KL} \tilde{M}_{LJ}.
\label{MMM1}
\end{align} 
By multiplying $\tilde{M}_{JK}$ to (\ref{MM}),
we find 
\begin{align}
(\tilde{M}_{KL} \tilde{M}_{LK}) \tilde{M}_{IJ}
= 2 \tilde{M}_{IK} \tilde{M}_{KL} \tilde{M}_{LJ}.
\label{MMM2}
\end{align}
Comparing (\ref{MMM1}) and (\ref{MMM2}), we obtain
\begin{align}
(\tilde{M}_{KL} \tilde{M}_{LK}) \tilde{M}_{IJ} = 0 \label{MMM3},\\
\tilde{M}_{IK} \tilde{M}_{KL} \tilde{M}_{LJ} = 0 \label{MMM4},
\end{align}
for generic masses.
If we impose the former constraint (\ref{MMM3}), 
the latter (\ref{MMM4}) can be derived from
(\ref{MMM2}), which originates in (\ref{MM0}).
Thus, independent constraints are (\ref{MM0}) and (\ref{MMM3}).

  In summary, the non-linear constraints for the meson operator are 
    \begin{align}
    \varepsilon^{IJKLM} \tilde{M}_{JK} \tilde{M}_{LM}
     =     0, 
           \\ 
    ({\rm Tr} \tilde{M}^2) \tilde{M}_{IJ}
     =     0.
    \end{align}
  These are equivalent constraints as (\ref{sum1}) and (\ref{sum2}).
  This matching of the non-linear constraints also indicates matching of the classical moduli space.
  This can be strong evidence of the Seiberg duality.

\section{Conclusion and discussion}
  In this paper, we have proposed a large number of new Seiberg dualities of $\CN=1$ generalized quiver gauge theories, 
  which originate in the S-dualities of $\CN=2$ superconformal gauge theories proposed by \cite{Gaiotto}.
  By deforming $\CN=2$ S-dual theories with adjoint mass terms, they flow to the Seiberg dual theories,
  where $\CN=1$ superconformal field theories are realized.
  We have shown some evidence for the existence of such infrared fixed points and the Seiberg dualities.
  We have found that for generic $SU(2)^n$ quiver gauge theories,
  the numbers of the exactly marginal operators are $2n$, which are universal for the proposed Seiberg dual theories.
  We have checked that the 't Hooft anomaly matching also hold for the Seiberg dual theories.
  As a simple example, we have considered the generalized Klebanov-Witten theory and its dual theory
  and demonstrated that chiral operators match between these two theories.
  We have also shown the matching of non-linear constraints for meson operators.
  
  In this paper, we have concentrated on the generalized $SU(2)^p$ quiver gauge theories.
  However, it is very interesting problem to generalize the gauge group to generic $SU(N)$.
  It would be possible to deform them to $\CN=1$ (see \cite{TW} for related discussion) 
  and discuss new Seiberg dualities of $SU(N)$ quiver gauge theories.
  This generalization is quite a non-trivial task 
  because the S-dualities of generic $SU(N)$ gauge theory is totally different 
  from that of $SU(2)$ as discussed in \cite{Gaiotto}.
  For example, the S-dualities for $SU(3)$ quiver gauge theories are based on the Argyres-Seiberg duality \cite{AS},
  where the $E_6$ superconformal field theory appears, whose $SU(2)$ subgroup is gauged.
  Since the explicit Lagrangian descriptions for the S-dual theories are not known,
  corresponding deformation as our analysis is not straightforward.
  It would also be interesting to generalize to $\CN=1$ $SO$-$USp$ quiver gauge theories,
  by using the S-dualities in $\CN=2$ superconformal $SO$-$USp$ quiver theories analyzed in \cite{Tachikawa}.

  For the $SU(N)$ gauge group, there are ${\cal N}=1$ SCFTs \cite{ArDo,TeYa} which 
  are obtained by adding a superpotential ${\rm Tr} \Phi^N$ to the 
  Argyres-Douglas ${\cal N}=2$ SCFT, even if there are no flavors.
  We expect that non-trivial ${\cal N}=1$ SCFTs can also be obtained from
  the ${\cal N}=2$ SCFTs considered in \cite{Gaiotto}
  by adding a superpotential ${\rm Tr} \Phi^N$.
  It will be interesting to study properties of 
  this type of SCFT.

\section*{Acknowledgements}
  We would like to thank Yuji Tachikawa for useful comments.
  Discussions during the YITP workshop YITP-W-09-04 on 
  ``Development of Quantum Field Theory and String Theory'' were useful to complete this work.
  The research of K.~M.~is supported in part by JSPS Research Fellowships
  for Young Scientists. S.~T.~is partly supported by the Japan Ministry
  of Education, Culture, Sports, Science and Technology. 
  M.~T.~is supported by JSPS Grant-in-Aid for Creative Scientific Research, No. 19GS0219.

\appendix

\section*{Appendix}
\section{Global symmetry of $\hat{A}_1$ theory}
\label{sec:USp4}
  A gauge theory corresponding to the left quiver diagram of Fig.~\ref{fig:KW} is known as 
  the $SU(2) \times SU(2)$ $\hat{A}_1$ theory \cite{DM, Cachazo:2001sg}.
  We can generalize it to the $SU(N) \times SU(N)$ $\hat{A}_1$ theory.
  This $SU(N)\times SU(N)$ theory is precisely the theory on $N$ D3 branes which probe the singular point of
  $\mathbb{C}^2 / \mathbb{Z}_2 \times \mathbb{C}$. 
  The $SU(N)\times SU(N)$   $\hat{A}_1$ theory possesses a global symmetry $SU(2)\times SU(2)$ 
  which rotates four bifundamental matters
  $(\mathbf{ N},\bar{\mathbf{ N}})$ and $(\bar{\mathbf{ N}},\mathbf{ N})$ respectively.
 
  In this appendix, we show that the global symmetry of $\mathcal{N}=2$ $\hat{A}_1$ theory is enhanced 
  to $USp(4)$ when we set $N=2$. 
  Since this special case is our interest, the enhanced flavor symmetry is important for our discussion.
   
Let us introduce vectors of the chiral fields $A_1$, $A_2$, $B_1$ and $B_2$ which 
are in the bifundamental representation $(\mathbf{ 2},\bar{\mathbf{ 2}})\simeq (\mathbf{ 2},\mathbf{ 2})$
\begin{align}
{A}=\left(\begin{array}{c}A_1 \\A_2\end{array}\right),\quad
{B}=\left(\begin{array}{c}B_1 \\B_2\end{array}\right).
\end{align}
Here we omit indices of gauge groups.
Then the superpotential of $SU(2)\times SU(2)$  $\hat{A}_1$ theory is 
given by
\begin{align}
W=\epsilon_{\dot{\alpha} \dot{\beta}} 
{}^{T}\!\! {A}^{ \alpha \dot{\alpha}}\cdot {B}^{\dot{\beta} \beta }\phi_{\alpha \beta}
+\epsilon_{\alpha \beta}
{}^{T}\!\! {B}^{\dot{\beta} \beta }\cdot{A}^{\alpha \dot{\alpha} }\tilde{\phi}_{\dot{\alpha} \dot{\beta} }.
\end{align}
Here $\alpha$ and $ \dot{\alpha}$ are the indices of the former and the latter $SU(2)$ gauge factor.
It is easy to see that the adjoint fields $\phi_{\alpha \beta}=\epsilon_{\beta \gamma}{\phi_{\alpha}}^{ \gamma}$ 
and $\tilde{\phi}_{\dot{\alpha} \dot{\beta} }=\epsilon_{\dot{\beta} \dot{\gamma} }
{{\tilde{\phi}_{\dot{\alpha} }}}^{\dot{\gamma}}$ are symmetric matrices.

Turning off the superpotential, the $SU(2)\times SU(2)$ Klebanov-Witten gauge theory has a global 
symmetry $U(4)$ which rotates the vector 
${}^{T}\!\! {Q}=({}^{T}\!\! {A},{}^{T}\!\! {B})$ as follows:
\begin{align}
{Q} \to U{Q}.
\end{align}
In this appendix, we study the subgroup of the unitary group $U(4)$ 
under which the superpotential is maintained  invariant. 
We represent an element of  $U(4)$ using $2\times 2$ matrices $a$, $b$, $c$, and $d$ as follows
\begin{align}
U=\left(\begin{array}{cc}a & b \\c & d\end{array}\right).
\end{align}
It acts on ${Q} $ as follows
\begin{align}
&{A}\to a{A}+b{B},\nonumber\\
&{B}\to c{A}+d{B}. \nonumber
\end{align}
Under the transformation, the first term of the superpotential becomes
\begin{align}
\epsilon_{\dot{\alpha} \dot{\beta}} 
{}^{T}\!\!{A}^{  \alpha \dot{\alpha} }\cdot{B}^{\dot{\beta} \beta }\phi_{\alpha \beta}
&\to
\epsilon_{\dot{\alpha} \dot{\beta}} 
({}^{T}\!\!{A}^{ \alpha \dot{\alpha}}  {}^{T}\! a+{}^{T}\!\!{B}^{ \dot{\alpha} \alpha}{}^{T}\! b)
\cdot
 (c{A}^{ \beta \dot{\beta} }+d{B}^{ \dot{\beta} \beta})
 \phi_{\alpha \beta}\nonumber\\ 
 &=\epsilon_{\dot{\alpha} \dot{\beta}} 
{}^{T}\!\!{A}^{ \alpha \dot{\alpha}}({}^T\! ad-{}^T\! bc){B}^{ \dot{\beta} \beta } \phi_{\alpha \beta}
+ \epsilon_{\dot{\alpha} \dot{\beta}} 
{}^{T}\!\!{A}^{ \alpha \dot{\alpha}}({}^T\!ac){A}^{  \beta \dot{\beta} }
\phi_{\alpha \beta}\nonumber\\
&\qquad\qquad\qquad\qquad\qquad\qquad\qquad\qquad\quad
+\epsilon_{\dot{\alpha} \dot{\beta}} {}^{T}\!\!{B}^{\dot{\alpha} \alpha}({}^{T}\!bd)
{B}^{ \dot{\beta} \beta } \phi_{\alpha \beta}.\nonumber
\end{align}
The invariance of the term implies the following constraints on the matrices
\begin{align}
{}^T \! ad-{}^T\!  bc=1,\quad {}^T \! ac={}^T \! ca, \quad {}^{T}\! bd={}^{T}\!\!  db.
\end{align}
Notice that the property $\phi_{\alpha \beta}=\phi_{\beta \alpha}$ implies
\begin{align}
\epsilon_{\dot{\alpha} \dot{\beta}} 
{}^{T}\!\!{A}^{ \alpha \dot{\alpha}}M{A}^{\beta \dot{\beta} } \phi_{\alpha \beta}
=
-\epsilon_{\dot{\alpha} \dot{\beta}} 
{}^{T}\!\!{A}^{  \alpha \dot{\alpha}}{}^T\!\! M{A}^{\beta \dot{\beta} } \phi_{\alpha \beta},
\end{align}
for a general element $M$ of $GL(2,\mathbb{C})$. 
The same relation holds for $B$.

The second term 
$\epsilon_{\alpha \beta}
{}^{T}\!\!{B}^{\dot{\beta} \beta }
\cdot{A}^{\alpha \dot{\alpha} }\tilde{\phi}_{\dot{\alpha} \dot{\beta} }$
 of the superpotential gives precisely same constraints.

Thus the superpotential maintains the subgroup of $U(4)$ whose elements satisfy the following relation of $4\times 4$ unitary matrix
\begin{align}
 \left(\begin{array}{cc}a & c \\b & d\end{array}\right)\left(\begin{array}{cc}0 & 1 \\-1 & 0\end{array}\right)\left(\begin{array}{cc}a & b \\c & d\end{array}\right)
=\left(\begin{array}{cc}0 & 1 \\-1 & 0\end{array}\right).
\end{align}
It means that the global symmetry of  $SU(2)\times SU(2)$ $\hat{A}_1$ theory is $USp(4)$.

\section{Mass deformation of $\hat{A}_1$ theory}
\label{sec:massdeformation}
In this section, we analyze the generalization of the Klebanov-Witten theory 
with a general marginal superpotential as an electric side of the duality.
We can generalize the $SU(N)\times SU(N)$ Klebanov-Witten theory by adding the following superpotential
\begin{align}
W=h^{\alpha \beta \dot{\alpha} \dot{\beta}} \mathrm{Tr}(A_\alpha B_{\dot{\alpha}} A_\beta B_{\dot{\beta}}).
\end{align}
Here $A$ and $B$ transform as $(\mathbf{ 2},\mathbf{ 1})$ and $ (\mathbf{ 1},\mathbf{ 2})$ under the global $SU(2)\times SU(2)$.
The trace operation vanishes some terms of $W$.
 Then the remaining superpotential consists of ten terms $(\mathbf{ 1},\mathbf{ 1})\oplus(\mathbf{ 3},\mathbf{ 3})$
\begin{align}
W=h\epsilon^{\alpha \beta}\epsilon^{ \dot{\alpha} \dot{\beta}} \mathrm{Tr}(A_\alpha B_{\dot{\alpha}} A_\beta B_{\dot{\beta}})
+h^{\{\alpha \beta \} \{\dot{\alpha} \dot{\beta}\}} \mathrm{Tr}(A_\alpha B_{\dot{\alpha}} A_\beta B_{\dot{\beta}}).
\end{align} 
Thus this is the general form of the superpotential by symmetry argument. 
In this subsection, we specialize it to $N=2$ and interpret it as a deformation of 
 $\hat{A}_1$ theory.

  Since we consider the gauge group $SU(2)\times SU(2)$ throughout the paper, 
  it is very convenient to rewrite it in $SO(4)$ notation. 
  Let us introduce six generators $\mathcal{I}^{i=1,2,3}$, $\mathcal{J}^{j=1,2,3}$
  of the gauge group $SU(2)\times SU(2)\simeq SO(4)$ 
    \bea
    \mathcal{I}^{i}
     =     (\sigma^2\otimes 1, \sigma^3\otimes \sigma^2, \sigma^1\otimes \sigma^2), 
           ~~~
    \mathcal{J}^{j}
     =     ( 1\otimes \sigma^2, \sigma^2\otimes \sigma^3, \sigma^2\otimes \sigma^1)
    \end{eqnarray}
  We expand the adjoint field by using Pauli matrices as basis
\begin{align}
{\phi^{a}}_b=
v^i {{\sigma^i}^{a}}_b,
\quad {{\tilde{\phi}}^{\dot{a}}}{}_{\dot{b}}
=
w^j {{\sigma^j}^{\dot{a}}}_{\dot{b}}.
\end{align}
Then we can collect the $SU(2)\subset SO(4)$ adjoint chiral fields $\phi$ and $\tilde{\phi}$ in $SO(4)$ (anti)self-dual matrices $V$, $W$ as follows: 
\begin{align}
V =\mathcal{I}^{i}v^i
=
\left(\begin{array}{cccc}
0 & -v^1 & -v^2 & -v^3 
\\v^1 & 0 & -v^3 & v^2
\\v^2 & v^3 & 0 & -v^1 
\\v^3 & -v^2 & v^1 & 0
\end{array}\right)
,\quad
W=\mathcal{J}^{j}w^j
=
\left(\begin{array}{cccc}
0 & -w^2 & -w^1 & -w^3 
\\w^2 & 0 & w^3 & -w^1
\\w^1 & -w^3 & 0 &-w^2 
\\w^3 & -w^1 & w^2 & 0
\end{array}\right).
\end{align}
It is easy to see that  these are $4\times 4$ self-dual and antiself-dual antisymmetric matrices. 
We can also represent an antisymmetric part of ${{A_1}^a}_{\dot{a}}{{B_1}^{\dot{a}}}_b+{{A_2}^a}_{\dot{a}} {{B_2}^{\dot{a}}}_b$
 and 
${{B_1}^{\dot{b}}}_{a}{{A_1}^{a}}_{\dot{a}}+{{B_2}^{\dot{b}}}_{a}{{A_2}^{a}}_{\dot{a}}$
 as self-dual and antiself-dual part of $4\times 4$ antisymmetric matrix $X$.
 The matrix is  given by $X_{a\dot{a}b\dot{b}} =
\epsilon_{\dot{a}\dot{b}}({{A_1}}_{\{a|\dot{c}|}{{B_1}^{\dot{c}}}_{b\}}+{{A_2}}_{\{a|\dot{c}|} {{B_2}^{\dot{c}}}_{b\}})
+\epsilon_{ab}
({{A_1}_{c\{\dot{a}}}{{B_1}_{\dot{b}\}}}^{c}+{{A_2}}_{c\{\dot{a}} {{B_2}_{\dot{b}\}}}^c)
$ in spinor indices. We rewrite it by lowering the spinor indices of fields as
\begin{align}
\epsilon_{\dot{a}\dot{b}}A_{\{a|\dot{c}|}{B^{\dot{c}}}_{b\}}
+\epsilon_{ab}{A_{c\{\dot{a}}}{B_{\dot{b}\}}}^{c}
&=-A_{\{a[\dot{a}}B_{\dot{b}] b\}}-A_{[a \{\dot{a}}B_{\dot{b}\} b]}\nonumber\\
&=-2(A_{a\dot{a}}B_{\dot{b}b}-A_{b\dot{b}}B_{\dot{a}a})\nonumber\\
&=-2(A_{\mu}B_{\nu}-A_{\nu}B_{\mu}).
\end{align}
Using this relation, we can write the matrix $X$ explicitly by using the index of $SO(4)$ vector
\begin{align}
X_{\mu\nu}=-2(A_{1[\mu}B_{1\nu]}+A_{2[\mu}B_{2\nu]}).
\end{align}
When we rename the fields as 
\begin{align}
(A_1, A_2, B_1, B_2) \to (Q^1, Q^2, Q^3, Q^4),
\end{align}
$X$ can be rewritten as 
\begin{align}
X_{\mu\nu} = - 2 J_{ij} Q_{\mu}{}^i Q_{\nu}^j, \label{def_X}
\end{align}
where $i,j=1,\cdots,4$ are indices of the global $USp(4)$ symmetry.

In this $SO(4)$ notation, the superpotential for $\hat{A}_1$ theory is
\begin{align}
W^{\hat{A}_1}
= \frac{1}{4} \tr (VX+WX).
\end{align}
Here $\tr$ is the trace over $4\times 4$ $SO(4)$ indices $\mu,\nu=1,\cdots,4$. 
Then we can deform the theory by adding an $\mathcal{N}=1$ general mass term
\begin{align}
W=W^{\hat{A}_1} - \frac{1}{4}(m_1\tr V^2+m_2\tr W^2)
\end{align}
At low energy, the theory flows to an $\mathcal{N} =1$ superconformal fixed point, as discussed above. 
The infrared theory is obtained by integrating out the massive chiral fields $V$, $W$.
\begin{align}
W^{\textrm{gKW}}&=
\frac{1}{16m_1}\tr ( X\cdot P^{\textrm{SD}}X)
+\frac{1}{16m_2}\tr ( X\cdot P^{\textrm{ASD}}X)\nonumber\\
&=\frac{1}{32}\left( \frac{1}{m_1} +\frac{1}{m_2}\right)
\tr ( X\cdot(P^{\textrm{SD}}+P^{\textrm{ASD}})X)\nonumber\\
&\hspace{5cm}+\frac{1}{32}\left( \frac{1}{m_1} -\frac{1}{m_2}\right)
\tr ( X\cdot (P^{\textrm{SD}}-P^{\textrm{ASD}})X).
\end{align}
where $P^{\textrm{(A)SD}}$ is the projector onto the (anti)self-dual part.
\begin{align}
&P^{\textrm{SD}}_{\mu\nu\rho\sigma}+P^{\textrm{ASD}}_{\mu\nu\rho\sigma}
=\frac{1}{2}(\delta_{\mu\rho}\delta_{\nu\sigma}-\delta_{\mu\sigma}\delta_{\nu\rho}),\nonumber\\
&P^{\textrm{SD}}_{\mu\nu\rho\sigma}-P^{\textrm{ASD}}_{\mu\nu\rho\sigma}
=\frac{1}{2}\epsilon_{\mu\nu\rho\sigma}.\nonumber
\end{align}
By substituting these explicit forms of the projectors, 
we obtain the superpotential of the generalized Klebanov-Witten theory
\begin{align}
\label{wgkw}
W^{\textrm{gKW}}=
\frac{1}{32} \left( \frac{1}{m_1} +\frac{1}{m_2}\right)
 X_{\mu\nu}X_{\mu\nu}
+ \frac{1}{64}\left( \frac{1}{m_1} -\frac{1}{m_2}\right)
\epsilon_{\mu\nu\rho\sigma}  X_{\mu\nu}X_{\rho\sigma}.
\end{align}
Here the first term is  the superpotential of the Klebanov-Witten theory. 
This term preserves the $SU(4)$ global symmetry because of their determinant representation in the next subsection. 
A generic mass deformation induces the second term which maintains only the original $USp(4)$ symmetry. 
By substituting (\ref{def_X}) into (\ref{wgkw}), we obtain the superpotential (\ref{W_KW}) 
appeared in section \ref{subsec:generalizedKW}.

\section{Chiral operators in the dual theory}
\label{sec:chiraloperators}
In this appendix, we identify
the independent operators in the chiral ring of the 
dual theory of generalized Klebanov-Witten theory.
Generally speaking,
when we consider chiral rings of a supersymmetric gauge theory,
we have to take into account that chiral ring relations for 
the gauge field strength 
$W_{\alpha} \propto \bar{D}_{\dot{\alpha}} \bar{D}^{\dot{\alpha}} [e^{-V} D_{\alpha} e^V]$ and other 
chiral superfields $\phi$ with arbitrary representation $R$ are given by
\begin{align}
W_{\alpha}{}^A (T_R{}^A)^a{}_b \phi^b
\propto \bar{D}_{\dot{\alpha}} \bar{D}^{\dot{\alpha}} \left[ e^{-V} D_{\alpha} (e^V \phi) \right]
\sim 0.
\label{ring_w}
\end{align}
Here, $D_{\alpha}, \bar{D}_{\dot{\alpha}}$ is a super covariant derivative,
$V$ is a vector superfield,
and $T_{R}$ is a generator of the gauge group in the representation $R$. 
This relation is also applicable for a product gauge group,
in which case $T_R{}^A$ should be replaced by generators of 
the product gauge group $T_{r_1}{}^{A_1} \otimes {\bf 1}$ 
and ${\bf 1} \otimes T_{r_2}{}^{A_2}$.

Writing the relation (\ref{ring_w}) explicitly for each field
$P$, $q$, $W_{\alpha}$, $w_{\alpha}$ in our model by using the discussion above, 
we obtain
\begin{align}
&(w_{\alpha})^{\dot{a}}{}_{\dot{b}} q^{\dot{b}} {}_I \sim 0 \label{ring1},\\
&(W_{\alpha})^a{}_b P^{b}{}_c{}^{\dot{a}} 
- P^{a}{}_b{}^{\dot{a}} (W_{\alpha})^b{}_c 
+ (w_{\alpha})^{\dot{a}}{}_{\dot{b}} P^{a}{}_c{}^{\dot{b}} \sim 0 
\label{ring2}, \\
&\{ W_{\alpha}, W_{\beta} \} \sim 0 \label{ring3}, \\
&\{ w_{\alpha}, w_{\beta} \} \sim 0 \label{ring4}.
\end{align}

First, we consider the operators invariant under the first gauge group $SU(2)_1$ but
not necessarily invariant under the second gauge group $SU(2)_2$.
The fields $P$ and $W_{\alpha}$ are in the adjoint representation of the first group.
A product of three adjoint fields $X=\sigma^i X^i$, $Y=\sigma^i Y^i$, $Z=\sigma^i Z^i$ 
of $SU(2)_1$ gauge group can be rewritten as 
\begin{align}
X^a{}_b Y^b{}_c Z^c{}_d  
= -i ( \varepsilon _{ijk} X^i Y^j Z^k ) \, \delta^a{}_d
+ {\rm Tr} (YZ) X^a{}_d - {\rm Tr} (ZX)  Y^a{}_d + {\rm Tr} (XY)  Z^a{}_d,
\label{3product}
\end{align} 
  where ``Tr'' is the trace of $SU(2)_1$.
  It indicates that a trace operator with more than three adjoint fields decomposes. 
Thus, candidates of the independent chiral operators which are invariant under 
the first gauge group $SU(2)_1$ are the trace operator with 
two or three fields because trace of a single adjoint field vanishes.
In the following, we discuss that trace operators with three adjoint fields
$P$ or $W$ vanish or reduce to trace operators with two adjoint superfields.
When we apply the equality (\ref{3product}) for the field $P^a{}_b{}^{\dot{a}}$,
the first term vanishes because the remaining index run only $\dot{a}=1,2$,
and thus, ${\rm Tr} P^3$ decomposes into ${\rm Tr} P^2 \times {\rm Tr} P$ and vanishes.
When we multiply $P^c{}_a{}^{\dot{c}}$ to the chiral ring relation (\ref{ring2}),
and taking into account the symmetry of the indices, we obtain
\begin{align}
2 (W_{\alpha})^a{}_b P^{b}{}_c{}^{\dot{a}} P^{c}{}_a{}^{\dot{c}}
= - (w_{\alpha})^{\dot{a}}{}_{\dot{b}} P^{a}{}_c{}^{\dot{b}} P^{c}{}_a{}^{\dot{c}}.
\label{wpp}
\end{align}
Thus, ${\rm Tr} WP^2$ reduces to ${\rm Tr} P^2$.
By using (\ref{ring3}) and by taking into account the symmetry of the $SU(2)_1$ indices,
we obtain
\begin{align}
(W_{\alpha})^a{}_b (W_{\beta})^b{}_c
\sim \frac{1}{4} \varepsilon_{\alpha\beta}  \delta^a{}_b
({\rm Tr} W^{\gamma} W_{\gamma}) .
\label{ww}
\end{align}
Thus, ${\rm Tr}W^2 P$ decomposes to the product of the glueball 
${\rm Tr}W^2$ and ${\rm Tr}P$, which vanishes.
Similarly, ${\rm Tr}W^3$ decomposes to the product of
${\rm Tr}W^2$ and ${\rm Tr}W$.

From the discussion above, the independent   
chiral operators invariant under the first gauge group $SU(2)_1$
are the following three operators;
\begin{align}
({\rm Tr} P^2)^{\dot{a}\dot{b}} = P^a{}_b{}^{\dot{a}} P^b{}_a{}^{\dot{b}}, \quad 
({\rm Tr} PW_{\alpha})^{\dot{a}}= P^a{}_b{}^{\dot{a}} (W_{\alpha})^b{}_a, 
\quad {\rm Tr} W_{\alpha} W_{\beta} = (W_{\alpha})^b{}_a(W_{\beta})^b{}_a.
\label{comb}
\end{align}
By using (\ref{ww}), we notice that the third operator can be rewritten as
\begin{align}
{\rm Tr} W_{\alpha} W_{\beta} 
= \frac{1}{2} \varepsilon_{\alpha\beta} {\rm Tr} W^{\gamma} W_{\gamma},
\end{align}
which is the glueball superfield.

Next, we consider the operators also invariant under 
the second gauge group $SU(2)_2$ by 
combining the fields $q$, $w$, and the operators in (\ref{comb}). 
Here, $(w_{\alpha})^{\dot{a}}{}_{\dot{b}}$ and 
$({\rm Tr}P^2)^{\dot{a}}{}_{\dot{c}} 
= ({\rm Tr}P^2)^{\dot{a}\dot{b}} \varepsilon_{\dot{b}\dot{c}}$ 
are in the adjoint representation 
while $q^{\dot{a}}{}_I$ and $({\rm Tr}PW_{\alpha})^{\dot{a}}$ are fundamental representation.
Gauge invariant operators are either ``loop type operators'', 
which are trace operators of adjoint superfields,
or ``linear type operators'', which are made up of
several adjoint superfields with two fundamental superfields at both end points.

We begin with the loop type operators.
As the chiral ring relation (\ref{ww}) is also applicable for the 
field strength $w_{\alpha}$ of the $SU(2)_2$ gauge group,
the operators with more than or equal to two field strength $w_{\alpha}$
is only the glueball superfield ${\rm tr} w^{\alpha} w_{\alpha}$.

In general, the square of adjoint fields $X=\sigma^i X^i$ 
of a $SU(2)$ gauge group can be rewritten as 
\begin{align}
X^{\dot{a}}{}_{\dot{b}} X^{\dot{b}}{}_{\dot{c}} 
= {\rm tr} (X^2) \delta^{\dot{a}}{}_{\dot{c}} .
\label{2product}
\end{align} 
It indicates that when we contract one set of the indices of two
$X=({\rm Tr}P^2)^{\dot{a}}{}_{\dot{b}}$,
the other set of indices are also contracted,
which results in the operator
\begin{align}
P^4 \equiv ({\rm Tr}P^2)^{\dot{a}}{}_{\dot{b}} ({\rm Tr}P^2)^{\dot{b}}{}_{\dot{a}}.
\label{P4}
\end{align}
Thus, the independent chiral operator 
more than or equal to two $({\rm Tr}P^2)^{\dot{a}}{}_{\dot{b}}$
is only this $P^4$.

From the discussion above,
we find that the independent gauge invariant chiral operators include at most two 
$({\rm Tr} P^2)^{\dot{a}}{}_{\dot{b}}$ and 
$(w_{\alpha})^{\dot{a}}{}_{\dot{b}}$ in total.
The candidates of loop type operators are as follows:
\begin{align} 
{\rm tr} \left( ww \right), \quad
{\rm tr} \left( w({\rm Tr}P^2) \right), \quad
{\rm tr} \left( ({\rm Tr}P^2)({\rm Tr}P^2) \right),
\label{tr_op}
\end{align}
where ``tr'' is the trace of the second gauge group $SU(2)_2$.

We go on to the linear type operators.
When they include field strength $w_{\alpha}$,
one of the gauge indices of $w_{\alpha}$ must be contracted to 
a combination of fields which is in the fundamental representation as a whole.
Taking into account that (\ref{ring_w}) is available also for 
a composite operator $\phi$, we notice that such operators vanish in the chiral ring.
Together with the discussion just below (\ref{2product}),
we notice that the linear type operators include at most one ${\rm Tr}P^2$
and two operators in the fundamental representation.
Thus, the candidates of the linear type operators are as follows:
\begin{align}
(q_I) (q_J) , \qquad
(q_I) ({\rm Tr} P^2) (q_J) , \qquad
(q_I) ({\rm Tr} PW_{\alpha}) , \qquad 
(q_I) ({\rm Tr} P^2) ({\rm Tr} PW_{\alpha}), \nonumber \\
({\rm Tr} PW_{\alpha}) ({\rm Tr} PW_{\beta}) , \qquad
({\rm Tr} PW_{\alpha}) ({\rm Tr} P^2) ({\rm Tr} PW_{\beta}), 
\label{chain_candidate}
\end{align}
where the gauge indices of $SU(2)_2$ are contracted properly.

In the following, we show that the last two operators in (\ref{chain_candidate})
actually decompose or vanish in the classical chiral ring.
By multiplying $W^d{}_a$ to (\ref{3product}), and by symmetrizing
$Y$ and $Z$, we show that
\begin{align}
{\rm Tr} (XYZW) + {\rm Tr} (XZYW) = 2 {\rm Tr} (YZ) \times {\rm Tr}(WX) 
\end{align}
Applying this identity to the fifth operator in (\ref{chain_candidate}), we obtain
\begin{align}
({\rm Tr} PW_{\alpha})^{\dot{a}} \varepsilon_{\dot{a}\dot{b}}
({\rm Tr}PW_{\beta})^{\dot{b}}   
= - \frac{1}{2} \varepsilon_{\dot{a}\dot{b}} 
\left( ({\rm Tr} PW_{\alpha}PW_{\beta}) ^{\dot{a}\dot{b}}
+ ({\rm Tr} P^2 W_{\alpha}W_{\beta})^{\dot{a}\dot{b}}  \right). 
\end{align}
By using (\ref{ring2}) to the first term of the right hand side 
of this equality, we obtain
\begin{align}
({\rm Tr} PW_{\alpha})^{\dot{a}} \varepsilon_{\dot{a}\dot{b}}
({\rm Tr}PW_{\beta}) ^{\dot{b}}  
= - \varepsilon_{\dot{a}\dot{b}} ({\rm Tr} P^2W_{\alpha}W_{\beta}) ^{\dot{a}\dot{b}}
+ \frac{1}{2} \varepsilon_{\dot{a}\dot{b}} 
(w_{\alpha})^{\dot{b}}{}_{\dot{c}} ({\rm Tr} P^2W_{\beta})^{\dot{a}\dot{c}}.
\end{align}
The equation (\ref{ww}) indicates that
the first term of this equality is decomposed into the product 
of the glueball ${\rm Tr} (W^{\alpha}W_{\alpha})$ and 
$\varepsilon_{\dot{a}\dot{b}} ({\rm Tr} P^2)^{\dot{a}\dot{b}}$ ,
where $\varepsilon_{\dot{a}\dot{b}} ({\rm Tr} P^2)^{\dot{a}\dot{b}}$ 
actually vanishes taking into account the symmetry of the indices.
We notice that the second term also decompose into 
the glueball ${\rm Tr} (w^{\alpha}w_{\alpha})$ and 
$\varepsilon_{\dot{a}\dot{b}} ({\rm Tr} PP)^{\dot{a}\dot{b}}$
by using the chiral ring relation (\ref{wpp}).
Thus, the operator $({\rm Tr} PW_{\alpha}) ({\rm Tr} PW_{\beta})$ 
in (\ref{chain_candidate}) vanishes identically
in the classical chiral ring.
Parallel discussion is possible for the last operator in (\ref{chain_candidate}),
and we show that this operator decompose into the product of
$P^4$ and a linear combination of two kinds of glueball superfield
${\rm Tr} (w^{\alpha}w_{\alpha})$ and ${\rm Tr} (W^{\alpha}W_{\alpha})$

So far, we have not imposed the equations of motion:
\begin{align}
-2 \left( \frac{1}{m_1} + \frac{1}{m_2} \right) 
P^{b}{}_a{}^{\dot{b}} \varepsilon_{\dot{b} \dot{c}} 
P^{c}{}_d{}^{\dot{c}} P^{d}{}_c{}^{\dot{d}} \varepsilon_{\dot{d} \dot{a}}
- \frac{2}{m_2} 
q^{\dot{b}}_I \varepsilon_{\dot{b}\dot{a}} 
P^{b}{}_a{}^{\dot{c}} \varepsilon_{\dot{c}\dot{d}} q^{\dot{d}}_I = 0, \label{eom1_dual}\\
- \frac{2}{m_2} 
\varepsilon_{\dot{a}\dot{b}} P^{a}{}_b{}^{\dot{b}} 
P^{b}{}_a{}^{\dot{c}} \varepsilon_{\dot{c}\dot{d}} q^{\dot{d}}_I
- \frac{2}{m_2} 
\varepsilon_{\dot{a}\dot{b}} q^{\dot{b}}_J 
q^{\dot{c}}_J \varepsilon_{\dot{c}\dot{d}} q^{\dot{d}}_I = 0, \label{eom2_dual}
\end{align}
which we write again for convenience.
By using the first equation of motion (\ref{eom1_dual}), we show that
unless $m_1+m_2 = 0$,
the last operator $P^4$ in (\ref{tr_op}) is proportional to the 
the second operator $(q_I) ({\rm Tr} P^2) (q_I)$ in (\ref{chain_candidate}), 
whose flavor indices are contracted:
\begin{align}
P^4 = - \frac{m_1}{m_1+m_2} (q_I) ({\rm Tr} P^2) (q_I). \label{P4_qPPq}
\end{align}
The second equation of motion (\ref{eom2_dual}) indicates that 
the last operator $(q_I) ({\rm Tr} P^2) (q_J)$ in (\ref{chain_candidate})
decomposes to the product of meson operator $(q_I) (q_J)$:
\begin{align}
(q_I) ({\rm Tr} P^2) (q_J) = - M_{IK} M_{KJ}. \label{qPPq_MM}
\end{align}

From the discussion in this appendix, 
we find that the independent gauge invariant operators 
in the classical chiral ring are (\ref{M_IJ}) - (\ref{S_2}).



\begin{thebibliography}{99}

\bibitem{Seiberg}
  N.~Seiberg,
  ``Electric - magnetic duality in supersymmetric nonAbelian gauge theories,''
  Nucl.\ Phys.\  B {\bf 435}, 129 (1995)
  [arXiv:hep-th/9411149].
  
\bibitem{ISreview}
  K.~A.~Intriligator and N.~Seiberg,
  ``Lectures on supersymmetric gauge theories and electric-magnetic  duality,''
  Nucl.\ Phys.\ Proc.\ Suppl.\  {\bf 45BC}, 1 (1996)
  [arXiv:hep-th/9509066].
  
\bibitem{SW}
  N.~Seiberg and E.~Witten,
  ``Monopoles, duality and chiral symmetry breaking in N=2 supersymmetric
  QCD,''
  Nucl.\ Phys.\  B {\bf 431}, 484 (1994)
  [arXiv:hep-th/9408099].
  
\bibitem{APShapere}
  P.~C.~Argyres, M.~R.~Plesser and A.~D.~Shapere,
  ``The Coulomb phase of N=2 supersymmetric QCD,''
  Phys.\ Rev.\ Lett.\  {\bf 75}, 1699 (1995)
  [arXiv:hep-th/9505100].

\bibitem{LS}
  R.~G.~Leigh and M.~J.~Strassler,
  ``Exactly Marginal Operators And Duality In Four-Dimensional N=1
  Supersymmetric Gauge Theory,''
  Nucl.\ Phys.\  B {\bf 447}, 95 (1995)
  [arXiv:hep-th/9503121].
  
\bibitem{APS}
  P.~C.~Argyres, M.~R.~Plesser and N.~Seiberg,
  ``The Moduli Space of N=2 SUSY {QCD} and Duality in N=1 SUSY {QCD},''
  Nucl.\ Phys.\  B {\bf 471}, 159 (1996)
  [arXiv:hep-th/9603042].
  
\bibitem{HMS}
  T.~Hirayama, N.~Maekawa and S.~Sugimoto,
  ``Deformations of N = 2 dualities to N = 1 dualities in SU, SO and USp  gauge
  theories,''
  Prog.\ Theor.\ Phys.\  {\bf 99}, 843 (1998)
  [arXiv:hep-th/9705069].

\bibitem{AILS}
  P.~C.~Argyres, K.~A.~Intriligator, R.~G.~Leigh and M.~J.~Strassler,
  ``On inherited duality in N = 1 d = 4 supersymmetric gauge theories,''
  JHEP {\bf 0004}, 029 (2000)
  [arXiv:hep-th/9910250].

\bibitem{Gaiotto}
  D.~Gaiotto,
  ``N=2 dualities,''
  arXiv:0904.2715 [hep-th].
  
\bibitem{Gaiotto:2009gz}
  D.~Gaiotto and J.~Maldacena,
  ``The gravity duals of N=2 superconformal field theories,''
  arXiv:0904.4466 [hep-th].
  
\bibitem{Tachikawa}
  Y.~Tachikawa,
  ``Six-dimensional $D_N$ theory and four-dimensional SO-USp quivers,''
  arXiv:0905.4074 [hep-th].

\bibitem{BBT}
  F.~Benini, S.~Benvenuti and Y.~Tachikawa,
  ``Webs of five-branes and N=2 superconformal field theories,''
  arXiv:0906.0359 [hep-th].

\bibitem{AGT}
  L.~F.~Alday, D.~Gaiotto and Y.~Tachikawa,
  ``Liouville Correlation Functions from Four-dimensional Gauge Theories,''
  arXiv:0906.3219 [hep-th].

\bibitem{ND}
  D.~Nanopoulos and D.~Xie,
  ``N=2 SU Quiver with USP Ends or SU Ends with Antisymmetric Matter,''
  arXiv:0907.1651 [hep-th].
  
\bibitem{Wyllard}
  N.~Wyllard,
  ``$A_{N-1}$ conformal Toda field theory correlation functions from conformal
  N=2 SU(N) quiver gauge theories,''
  arXiv:0907.2189 [hep-th].
  
\bibitem{KW}
  I.~R.~Klebanov and E.~Witten,
  ``Superconformal field theory on threebranes at a Calabi-Yau  singularity,''
  Nucl.\ Phys.\  B {\bf 536}, 199 (1998)
  [arXiv:hep-th/9807080].

\bibitem{Cachazo:2001gh}
  F.~Cachazo, S.~Katz and C.~Vafa,
  ``Geometric transitions and N = 1 quiver theories,''
  arXiv:hep-th/0108120.

\bibitem{Cachazo:2001sg}
  F.~Cachazo, B.~Fiol, K.~A.~Intriligator, S.~Katz and C.~Vafa,
  ``A geometric unification of dualities,"
  Nucl.\ Phys.\  B {\bf 628}, 3 (2002)
  [arXiv:hep-th/0110028].

\bibitem{DM}
  M.~R.~Douglas and G.~W.~Moore,
  ``D-branes, Quivers, and ALE Instantons,''
  arXiv:hep-th/9603167.

\bibitem{IS}
  K.~A.~Intriligator and N.~Seiberg,
  ``Duality, monopoles, dyons, confinement and oblique confinement in
  supersymmetric SO(N(c)) gauge theories,''
  Nucl.\ Phys.\  B {\bf 444}, 125 (1995)
  [arXiv:hep-th/9503179].

\bibitem{AS}
  P.~C.~Argyres and N.~Seiberg,
  ``S-duality in N=2 supersymmetric gauge theories,''
  JHEP {\bf 0712}, 088 (2007)
  [arXiv:0711.0054 [hep-th]].

\bibitem{TW}
  Y.~Tachikawa and B.~Wecht,
  ``27/32,''
  arXiv:0906.0965 [hep-th];
  F.~Benini, Y.~Ookouchi, Y.~Tachikawa and B.~Wecht, 
  in preparation.

\bibitem{ArDo}
  P.~C.~Argyres and M.~R.~Douglas,
  ``New phenomena in SU(3) supersymmetric gauge theory,''
  Nucl.\ Phys.\  B {\bf 448} (1995) 93
  [arXiv:hep-th/9505062].

\bibitem{TeYa}
  S.~Terashima and S.~K.~Yang,
  ``Confining phase of N = 1 supersymmetric gauge theories and N = 2  massless
  solitons,''
  Phys.\ Lett.\  B {\bf 391} (1997) 107
  [arXiv:hep-th/9607151];
  ``ADE confining phase superpotentials,''
  Nucl.\ Phys.\  B {\bf 519} (1998) 453
  [arXiv:hep-th/9706076].

\end{thebibliography}
\end{document}